\documentclass[%
 reprint,
 superscriptaddress,
nofootinbib,
 amsmath,amssymb,
 aps,
 prd,
]{revtex4-2}
\usepackage{graphicx}
\usepackage{dcolumn}
\usepackage{bm}
\usepackage{hyperref}
\usepackage{xcolor}
\usepackage{subcaption} 
\newcommand{\si}[1]{\:\mathrm{#1}}
\newcommand{\ratio}{$N_{\mu} / N_{\mathrm{e}}$ }

\DeclareUnicodeCharacter{2212}{-}

\begin{document}
\title{Approach for composition measurement of cosmic rays using muon-to-electron ratio observed by LHAASO-KM2A}

\author{Xishui Tian}
 \affiliation{Department of Astronomy, School of Physics,
             Peking University, Beijing 100871, China}
\author{Zhuo Li}
 \email{zhuo.li@pku.edu.cn}
 \affiliation{Department of Astronomy, School of Physics,
             Peking University, Beijing 100871, China}
 \affiliation{Kavli Institute for Astronomy and Astrophysics, Peking University, Beijing 100871, China}      
\author{Quanbu Gou}
 \email{gouqb@ihep.ac.cn}
\affiliation{Key Laboratory of Particle Astrophysics,
Institute of High Energy Physics, Chinese Academy of Sciences, Beijing 100049, China}
\affiliation{University of Chinese Academy of Sciences, Beijing 100049, China}
\affiliation{Tianfu Cosmic Ray Research Center,  Chengdu 610000, Sichuan, China}

\author{Hengying Zhang}
\affiliation{School of Physics and Astronomy, Yunnan University, 650091 Kunming, Yunnan, China}
\author{Huihai He}
\affiliation{Key Laboratory of Particle Astrophysics,
Institute of High Energy Physics, Chinese Academy of Sciences, Beijing 100049, China}
\affiliation{University of Chinese Academy of Sciences, Beijing 100049, China}
\affiliation{Tianfu Cosmic Ray Research Center,  Chengdu 610000, Sichuan, China}
\author{Cunfeng Feng}
\affiliation{Key Laboratory of Particle Physics and Particle Irradiation (MOE), Institute of Frontier and Interdisciplinary Science, Shandong University, Qingdao, Shandong 266237, China}
\author{Giuseppe Di Sciascio}
\affiliation{INFN - Roma Tor Vergata, Via della Ricerca Scientifica 1, 00133 Rome, Italy}

\date{\today}
\begin{abstract}
    Composition measurement of cosmic rays (CRs) around the knee of the CR energy spectrum is crucial for studying the processes of particle acceleration and propagation of Galactic CRs.
    The Square Kilometer Array (KM2A) of Large High Altitude Air Shower Observatory (LHAASO) can provide precise measurement of the muonic and electromagnetic (em.) components in CR-induced extensive air showers, and hence a good chance to disentangle the CR composition. Here we propose an approach of decomposing CR compositions with the number ratio between muons and em. particles ($N_{{\mu}} / N_{\mathrm{e}}$) observed by LHAASO-KM2A: we reconstruct the energy spectra of individual CR compositions
    by fitting $N_{{\mu}} / N_{\mathrm{e}}$
    distributions in each reconstructed energy bin using the template shapes of $N_{{\mu}} / N_{\mathrm{e}}$ distributions of individual CR compositions based on Monte Carlo (MC) simulation. We evaluate the performance of this approach with MC tests where mock data of LHAASO-KM2A are generated by MC simulation. We show that the input composition model can be well recovered in this approach, {independent of the CR composition model adopted in the MC simulation for the template distributions.} The uncertainties of the reconstructed spectra at $<20\si{PeV}$, mainly limited by simulation statistics,
    are $\leq 7\%$ for proton, He, and Fe groups, and 
    $\leq 8\%$ and $\leq 16\%$ for CNO and MgAlSi groups, respectively.
\end{abstract}

\maketitle

\section{Introduction}
    Cosmic rays (CRs) are high energy
    atomic nuclei of astrophysical origins.
    The observed all-particle energy spectrum of CRs
    is approximately a power-law
    from $\si{GeV}$ to beyond $10^{11}\si{GeV}$ with
    several breaks where the indices of the power-law change.
    The energy spectrum of CRs steepens around {$ 4\si{PeV}$},
    with the power-law index changing from {$-2.7$} to {$-3.1$}  \cite{RN588}, which is the so-called ``knee" of the CR spectrum.
    The physical cause of the CR spectral knee is still unclear, although many possible reasons had been discussed.
    First, the break may arise 
    from the particle acceleration capability of Galactic CR accelerators \cite{RN576,RN577},
    i.e., the maximum energy
    of CRs that the accelerators can produce. 
    For example, supernova remnants (SNRs) are well discussed to be potential candidates of Galactic CR accelerators \cite{RN575}, which are believed to accelerate protons up to {$\sim 100 \si{TeV}$}
    by shock acceleration \cite{RN559}, but hardly up to PeV range  \cite{RN597}.
    Secondly, the knee may be caused
    by propagation effects \cite{RN585,RN582,RN578}.
    CRs propagate diffusively
    in the Galactic magnetic field (GMF) \cite{RN594,RN595}. The propagation process depends on particle energy. Low and high energy CRs follow different propagation patterns, which may lead to a spectral break in between. 
    
    Either particle acceleration or propagation process is expected to be governed by electromagnetic (em.) interaction with the background magnetic field, which is rigidity dependent, i.e., the particles with the same rigidity may undergo the similar acceleration and/or propagation processes. Rigidity-dependent
    spectra are expected for different types of nuclei  \cite{RN433,RN426}.
    Therefore, precise measurements of elemental composition around the knee
    are crucial to understanding the acceleration and propagation of CRs, and
    deciphering the mechanism accounting for the knee.
    
    
    Because of the low CR flux at high energies, CRs of $>100\si{TeV}$ are detected via extensive air showers by ground-based detectors.
    Due to the fluctuations in air showers, 
    it is hard to reconstruct the primary mass for single air shower event.
    However, elemental energy spectra grouped by CR mass
    can be statistically reconstructed from observables on the ground.
    Observationally, it is common to decompose the CR compositions at $>100\si{TeV}$ into five groups, following the direct measurements,  
    proton, He, CNO, MgAlSi, and Fe  \cite{RN557}.
   
    Many ground-based experiments had tried
    to measure the CR compositions around the knee region.
    But the results among experiments are inconsistent,
    in particular, on the breaks of proton and He spectra.
    {KASCADE} decomposed the primary CRs from $1\si{PeV}$ to $100\si{PeV}$
    into five mass groups,
    represented by proton, He, C, Si, and Fe,
    by unfolding the distribution of secondary electrons and muons \cite{RN544}.
    The reconstructed spectra of proton and He break beyond PeV,
    which indicates the knee of the all particle spectrum
    results from the breaks of proton and He, namely the light component.
    On the contrary, a light knee below PeV is observed by Tibet AS$\gamma$  \cite{RN546} and ARGO-YBJ plus Wide Field Cherenkov Telescope (WFCT)  \cite{RN427}
    measurements.
    Tibet AS$\gamma$ measured the energy spectra of proton
    and He from $1\si{PeV}$ to $10\si{PeV}$ using artificial
    neutral network to reconstruct the primary mass
    and select proton-like events and light-like events \cite{RN546}.
    The spectra of proton and He are expressed as $E^{-3}$ power-law
    from $1\si{PeV}$ to $10\si{PeV}$,
    suggesting the light component breaks below 1 PeV.
    ARGO-YBJ+WFCT measured the energy spectrum of
    proton plus He by selecting light-like events
    with a two-dimension cut applied to
    observables from surface detectors and the Cherenkov telescope.
    The spectrum of light nuclei measured by ARGO-YBJ+WFCT 
    breaks around {$700\si{TeV}$} \cite{RN427},
    which is compatible with the results from Tibet AS$\gamma$,
    but at odds with the KASCADE measurements.
    
   The inconsistency on the CR composition may
    come from the limitations of each experiment.
    For KASCADE, the detectors are deployed at the sea level (grammage $\sim 10^{3} \si{g\,cm^{-2}}$),
    which is far from the shower maximum of PeV energies ($X_{\rm max } \sim 600 \si{g\,cm^{-2}}$).
    Therefore, the measurement suffers
    more uncertainty from the atmospheric absorption.
    On the other hand, though located around the shower maximum of PeV showers,
    Tibet AS$\gamma$ and ARGO-YBJ lack good muon measurements in the shower,
    which is crucial to distinguishing different nuclei.
    Thus, an observatory at high altitude as well as having good
    measurements of secondary muons is necessary to settle 
    the question of the composition around the knee.

   The Large High Altitude Air Shower Observatory (LHAASO)  \cite{RN590},
    with an array area of $\sim 1.3 \si{km^{2}}$ 
    and located at Mountain Haizi, Daocheng, China of
    $4410\si{m}$ above sea level,
    is designed for precise measurements of high energy CR and gamma-ray induced air showers.
    It consists of three sub-arrays -- Square Kilometer Array (KM2A),
    Water Cherenkov Detector Array (WCDA),
    and Wide Field Cherenkov Telescope Array (WFCTA) -- and observes air showers from high energy CRs and gamma-rays using hybrid techniques \cite{RN590}. In one of the key goals for CR observations, LHAASO aims to measure the energy spectra, i.e., the spectral knee, of the light (proton or H+He) and Fe groups, 
    by using multiple observables from KM2A, WFCTA and WCDA data combined to select air shower events for individual groups with high purity \cite{RN623,RN622,RN624}.
    
    We note that LHAASO-KM2A can be very powerfal for CR composition measurement around the knee region. First, the high altitude of LHAASO, near the shower maximum for CR energy around $4\si{PeV}$ allows the energy reconstruction of primary CRs
    with less dependence on the CR composition \cite{RN542}.
    The insensitiveness of energy reconstruction to
    the primary mass will greatly facilitate composition
    studies. 
    Second, thanks to the dense arrangement of secondary detectors in LHAASO-KM2A,
    precise measurement of lateral distribution of secondaries in air showers
    is allowed, and low energy CRs can be observed down to $30\si{TeV}$.
    Third, the large area makes LHAASO-KM2A capable of observing CR with energies up to
    $100\si{PeV}$. 
    The wide energy range of CRs observed by LHAASO-KM2A
    will provide complementary measurements between direct
    measurements of low-energy CRs via satellites/balloons
    and high-energy CR observations by ground arrays.
    Forth, LHAASO-KM2A is equipped with 
    two types of detectors to probe both the muonic
    and em. components of the shower, which is crucial for mass
    separation in the composition reconstruction \cite{RN591}.
    
    In this work, we propose an approach of composition measurement by LHAASO-KM2A. Given the measurements of the muonic and em. components of air showers, we use the number ratio between muons and em. particles (\ratio) of the air shower as the mass indicator of primary CRs
    to reconstruct the elemental energy spectra.
    We carry out detection simulation to evaluate the performance of
    LHAASO-KM2A in reconstructing the elemental energy
    spectra. 
    The paper is organised as
    follows: Section \ref{sec:LHAASO} introduces 
    the detector layout of LHAASO-KM2A
    and the detection simulation used in this analysis.
    Section \ref{sec:method} describes the analysis procedure,
    including the energy reconstruction, the mass separation,
    and the reconstruction of elemental energy spectra.
    Section \ref{sec:result} demonstrates the capability
    of reconstructing the elemental energy spectra with
    Monte Carlo (MC) tests and presents the uncertainties in reconstruction
    shown in the MC tests.
    Section \ref{sec:summary} is summary and discussion on the approach.
    
\section{Experiment and Simulation \label{sec:LHAASO}}

  \subsection{LHAASO-KM2A layout}
        Hybrid techniques for air shower observations are used in LHAASO-KM2A, which consists of 5216 em. detectors (ED)
        and 1188 muon detectors (MD), independently measuring
        the secondary particles in the air shower.
        ED is scintillator detector with a sensitive area of $1\si{m^2}$.
        EDs are deployed over a circle of radius $635\si{m}$.
        The spacing between EDs is $15\si{m}$ in the central region
        (with radius $R < 575\ \si{m}$) and $30\ \si{m}$
        at outskirts ($575 < R < 635\ \si{m}$).
        The different spacing is used to better reconstruct the events
        for showers near the edge of the central array.
        MD is water Cherenkov detector that has an area
        of $36\si{m^{2}}$ buried under $2.5\si{m}$ soil to shield it
        from the electromagnetic particles in the shower.
        MDs are deployed in the central region of radius $575\si{m}$
        with a spacing of $30\si{m}$. Details about the design
        and performance of ED and MD can be found in  \cite{RN560}.
   
    \subsection{Detection simulation\label{subsec:mc}}
        In order to evaluate the KM2A performance in composition observations,
        we carry out MC simulation for the detection. We simulate the interactions
        and propagation of air showers
        in the atmosphere and the detector response on the ground.
        The simulated CR showers are reconstructed by
        utilising the hit information of the triggered detectors.

        In the simulation, five types of nuclei are used:
        proton, He, N (denoted by CNO), Al (denoted by MgAlSi), and Fe.
        The shower energy is sampled from a $E^{-2}$ power-law
        from $10\si{TeV}$ to {$50\si{PeV}$}.
        The incident directions are drawn from
        an isotropic distribution in the range
        of zenith $0\, \text{--}\, 40^{\circ}$
        and azimuth $0\,\text{--}\,360^{\circ}$.
        
        The simulation is produced by
        {\small CORSIKA} 7.7410 \cite{RN553}
        and {\small G4KM2A}  \cite{RN570}, for air show simulations and detector response, respectively.
        {\small CORSIKA} samples the generation of secondary particles
        in the atmosphere and the propagation of the shower to the ground.
        The selected interaction models are
        QGSJetII-04 \cite{RN563} for high energy ($>80 \si{GeV}$) interactions
        and {\small FLUKA} \cite{RN564} for low energy ($< 80 \si{GeV}$) interactions.
        The atmosphere model used for {\small CORSIKA} is the US standard atmosphere,
        with a grammage of $597\si{g\, cm^{-2}}$ for the LHAASO site.
        The magnetic field is set to $B_{x} = 34.6\si{\mu T}$ and
        $B_{z} = 36.1 \si{\mu T}$,
        where the x-axis points to the north
        and the z-axis points vertically downwards.

        In the detector response simulation with {\small G4KM2A},
        the shower cores of the injected showers are uniformly distributed
        in an annulus {$ 260 < R < 480$\,m} from the array center
        in order to boost sampling efficiency considering the current
        event selection condition (See Appendix \ref{app:cut}). 
        In {\small G4KM2A}, one {\small CORSIKA} shower is reused 20 times 
        to increase the statistics, i.e., a same shower is injected 20 times
        with different shower core positions.
        The total number of simulated showers of all five mass groups
        are around {$5.555 \times 10^{7}$} before trigger, 
        of which {$1.36 \times 10^{6}$} events survive after the trigger and event selections.
        
        The trigger events in the simulation are further being reconstructed. 
        The hit information after filtering the noise from triggered detectors 
        is translated into the properties of the primary particle.
        The shower core position and the incident direction is
        fitted by the hits from all triggered EDs.
        The time resolution of ED during reconstruction is set to $0.2\si{ns}$.
        from the charges collected by the EDs (MDs).
        
\section{CR decomposing method \label{sec:method}}
    Observations of CR compositions call for at least 
    two orthogonal
    measurements of an air shower to obtain primary energy and mass
    at the same time.
    High energy CR particle interacts with nuclei in the atmosphere
    and produces secondary mesons, which further
    decay or interact during propagation. Thus a high energy
    CR particle will initiate a particle cascade
    made up of decay and interaction in the atmosphere.
    Most of the  secondary particles that reach the ground 
    are em. particles, including electrons/positrons
    and photons, and muons. em. particles are mostly resulted from the
    decay of neutral pions, while muons are from the decay of
    charged pions. The secondary em. particles and
    muons are related to  the shower development
    which depends on the energy and mass of the primary CR particle.
    Therefore, taking advantage of the large and uniform coverage
    of EDs and MDs of LHAASO-KM2A, we can use the number of em. particles, $N_{\mathrm{e}}$,
    and the number of muons, $N_{\mathrm{\mu}}$, of each shower to estimate the CR energy and composition \footnote{Note, $N_{\mu}$ and $N_{\mathrm{e}}$ are the numbers of secondary particles detected by all triggered MDs and EDs located $40\,\text{--}\, 200\si{m}$ from the shower core on the shower plane, respectively.} and hence reconstruct the elemental energy spectra.

    The proposed method is as follows. First, we reconstruct the shower energies using a composition insensitive method, and then bin events by reconstructed energies. Next, taking advantages of the mass-dependent \ratio ratio, we decompose CR compositions by fitting the \ratio distribution in each reconstructed energy bin. See below for the details.
    
    \subsection{Mass-insensitive energy reconstruction}
        The primary CR energy can be given by summing up the two components of CR showers, i.e., em. and muonic components.
        The altitude of the LHAASO site is around the shower maximum for the knee-region CRs. 
        The proximity to the shower maximum improves the energy reconstruction, and weakens the dependence on the primary CR composition. An approach for energy reconstruction of LHAASO-KM2A has been given in Ref.  \cite{RN542}, which is only weakly dependent on the primary mass .   
        For this analysis, we use Eq 9 in Ref.  \cite{RN542} to reconstruct the primary CR energy $E$, i.e.,
        \begin{equation}
            \log\left( {E/\mathrm{GeV}}\right) = 2.768 + \log{(2.8 N_{\mu} + N_{\mathrm{e}})}.
        \end{equation}
        The resolution of energy reconstruction is better than $15\%$ above
        $300\si{TeV}$ and the energy bias is smaller than $5\%$ \cite{RN542}, independent of the primary mass.
        
    \subsection{Mass dependence of \ratio ratio}
        The primary mass affects the properties of air showers,
        and hence the observables on the ground.
        For CRs of the same energy,
        heavy nuclei tend to produce more muons and
        less em. particles than light nuclei
        at the shower maximum \cite{RN297}.
        Therefore, \ratio can be a good estimator of the primary mass, i.e., heavier nucleus
        shows larger \ratio around the shower maximum \cite{RN561}.
        
        Moreover, the shower generation and development
        has intrinsic fluctuations because the particle interaction is a stochastic process.
        The fluctuation is mostly resulted from the first few interactions of the shower. As a result, showers of the same CRs, with the same energy and composition, show a distribution
        of \ratio instead of a constant. Besides, the scatter of the fluctuation is also dependent on the primary mass, with proton showers being more fluctuating while
        those of heavy nuclei showing less fluctuation,
        due to the averaging effects of more nucleons.
        Therefore, the width of the \ratio distribution
        is determined by the primary composition.

        The \ratio distribution encapsulates the primary composition
        by both its shape and mean (or peak) value.
        \begin{figure}
            \centering
            \includegraphics[width=3.4in]{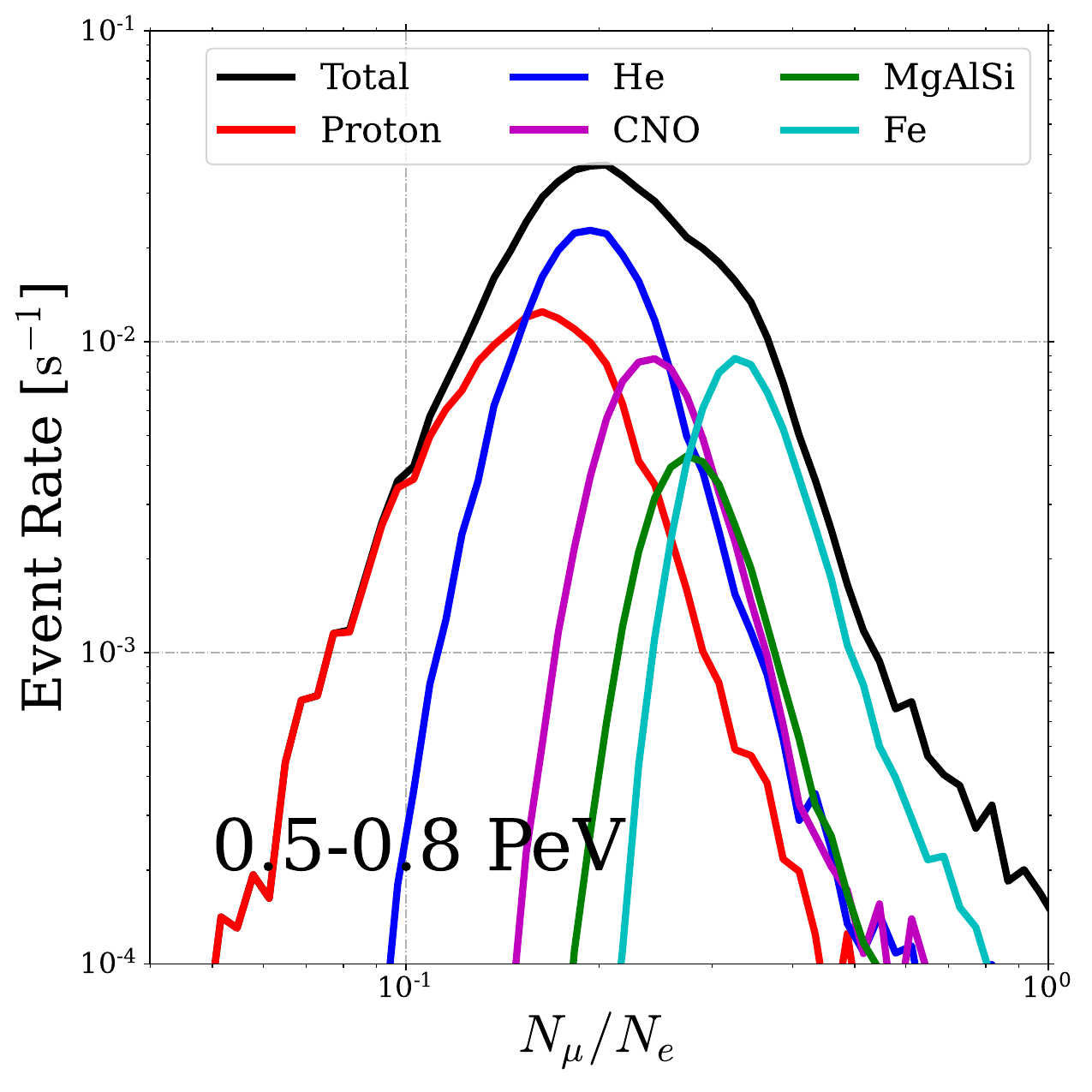}
            \caption{Simulated \ratio distribution in the reconstructed energy bin of $0.5\,\text{--}\, 0.8 \si{PeV}$, assuming the H3a model \cite{RN261}.
            The black line is the total distribution and the colorful
            lines are distributions of each mass groups
            as annotated in the figure legend.
            }
            \label{fig:H3a-ratio-distribution}
        \end{figure}
        Figure \ref{fig:H3a-ratio-distribution} shows the simulated \ratio distribution
        in the reconstructed energy bin of $0.5-0.8$ PeV, assuming a CR composition model, namely the H3a composition model \cite{RN261}, which is a three-spectral-component fit to observed CR spectra with rigidity-dependent cutoffs. The distributions of each mass group, along with the total distribution, have been shown. Despite of the fluctuation due to the low statistics in the simulation on the left and right ends of the distribution, it is clearly shown that the distributions of each mass group
        differ from each other in two aspects, corresponding to the shower development.
        First, the peaks of five mass groups are
        well separated. Statistically, the proton showers tend to
        have the smallest \ratio values, dominating the
        total distribution at the low-value end. The Fe showers have the largest \ratio ratio, 
        dominating the high-value end of the total distribution.
        The peak position reflects the most probable value of \ratio
        for each mass group, which results from the fact that
        heavy nucleus produces more muons than the light nucleus.
        Second, the width of the distributions is different between mass groups.
        The proton showers disperse over a wider range compared to heavier nuclei,
        whereas the Fe distribution is narrower, resulting from smaller fluctuations of the development of Fe shower.
        
        The fact that different mass groups are well separated in \ratio distribution allows for decomposing CR compositions and reconstructing
        the elemental energy spectra by fitting \ratio distribution, as discussed in the following.
        
    \subsection{{Derivation of fluxes for different mass groups}\label{sec:fitting}} 
        In our approach, we generate template \ratio distributions for individual mass groups by MC simulation, with which we decompose the observed \ratio distribution into different mass groups, and then reconstruct the elemental energy spectra. The events of both observed data and MC are binned in $E$ and ${N_{\mu}}/N_{\rm e}$. We fit the observed \ratio distribution independently in each energy bin with the template distributions. 
        The fluxes of different mass groups in a given energy bin are given by the best fit. 
        Therefore, the elemental energy spectra of CRs 
        are reconstructed by fitting the fluxes of the five mass groups in separate energy bins.

        Note, we use the template \ratio distributions of individual mass groups coming from MC simulation rather than doing any parameterization of the distribution.
        It is necessary to assume a {\em template model} for the spectral shape
        of each mass group within one energy bin. However, see discussion in section \ref{sec:discussion:template model}, the choice of the template model and hence the spectral shape
        could have little effect on the fitting results,
        because of the narrow binning of energies.
        Thus, the MC simulated templates can well describe the profile of the distribution for each mass group at given energy. Moreover, we also note that the fitting is done independently for energy bins, so that the method is CR composition model independent, avoiding circular reasoning in measuring CR compositions.
        
        Therefore, the free parameters in the fitting of the observed total distribution are only the fluxes of individual mass groups, here denoted as the normalization factors $f^k$
        with respect to the assumed flux for the each mass group,
        where $k$ stands for the nucleus types. For the best-fit of $f^k$, the flux of nucleus $k$ is derived to be $f^k F^{k}_0(E_i)$,
        with $F^{k}_0(E_i)$ being the assumed flux at energy $E_{i}$ when generating the template distributions.
        
        To obtain the best fit, we use Markov chain Monte Carlo (MCMC)
        implemented in Ref. \cite{RN572}
        to sample the distribution of likelihood.
        The best fit value is the median value
        of the \emph{a posteriori} distribution and the statistical uncertainty
        is bracketed by the $16\%$ and $84\%$ percentiles.
        Detailed fitting procedure
        is described in App. \ref{app:mcmc}.
        
\section{Monte Carlo Test\label{sec:result}}
    In the following, we carry out MC simulation to test the validity of the composition measurement approach proposed in section \ref{sec:method}, and derive the uncertainty of the reconstructed elementary spectra.
    The observational data is replaced by the mock data from MC simulation assuming a certain {\em input composition model}, thus the mock data will be fitted with the procedures described in Sec. \ref{sec:fitting}.
    The fitting results are compared with the input composition model
    of the mock data to examine the capability of recovering
    the CR compositions.
    In this analysis, we use the same simulation data set
    for producing the mock data and the MC templates.         
    
    In order to estimate the effects of template model and the input model on decomposing the CR composition,
    the method should be tested under various assumptions.
    Table \ref{tab:MCtests} summarizes the three MC tests in this work,
    showing the input and template composition models. 
    We consider three choices of CR models. Besides the H3a model \cite{RN261}, we take the Global Spline Fit model (GSF), which is the latest empirical fit from the available experiment data \cite{RN569}, and the Poly-gonato (PG) model \cite{RN423}, which is also empirical fit of observed data,
    assuming rigidity-dependence for spectral cutoffs of different nuclei.    
    We only consider the five leading mass groups for simplicity:
    proton, He, CNO, MgAlSi, and Fe.
    The simulation of CNO (MgAlSi) mass group is
    simplified by approximating the mass group
    by the nucleus in the middle, i.e., N (Al), whose
    flux is the summation of the three nuclei in the mass group.
    
    \begin{table}
    \caption{\label{tab:MCtests} Three MC tests.}
    \begin{ruledtabular}
    \begin{tabular}{ccc}
    Test & Input model & Template model  \\ \hline
    I   & GSF         & H3a               \\ 
    II   & GSF         & PG                 \\ 
    III   & PG          & H3a               \\ 
    \end{tabular}
    \end{ruledtabular}
    Note, template model is assumed to generate the template $N_\mu/N_e$ distributions of individual mass group, and input model to generate the mock observational data. 
    \end{table}

    \subsection{Benchmark test}
    For the benchmark test, Test I, the template CR composition model
    is taken to be H3a model for the template \ratio distributions of individual mass groups,
    and the input CR composition model is taken to be GSF model for the mock data distribution.
    The results of the benchmark test (Test I) are shown in this section, which is followed by the other tests with the input or template model changed to estimate the effects of the model assumptions.

    The events are binned in $E$ and \ratio in log scales. 
    The energy range of this analysis is $E=0.2\,\textup{--}\,20\si{PeV}$. This wide energy range helps to better probe the 
    composition evolution around the knee
    and the possible spectral breaks of elemental energy spectra.
    The energy is binned in log space, $\log (E/{\rm GeV})$,
    with a step size of {$0.2$}, in comparison with the energy resolution of better than $15\%$ above $300\si{TeV}$.
    The \ratio ratio
    is evenly binned in log space with a step size of {$0.025$}
    in the range of \ratio$=10^{-1.5}-10^{0.5}$,
    which is a compromise between the number of bins
    and the statistic within each bin.

    Presented below are the fitting results of \ratio distribution
    and the reconstructed elemental energy spectra and discussions on uncertainties.
    
    \subsubsection{Best fit of \ratio distribution \label{sec:result-ratio}}
\begin{figure}
            \centering
            \includegraphics[width=3.4in]{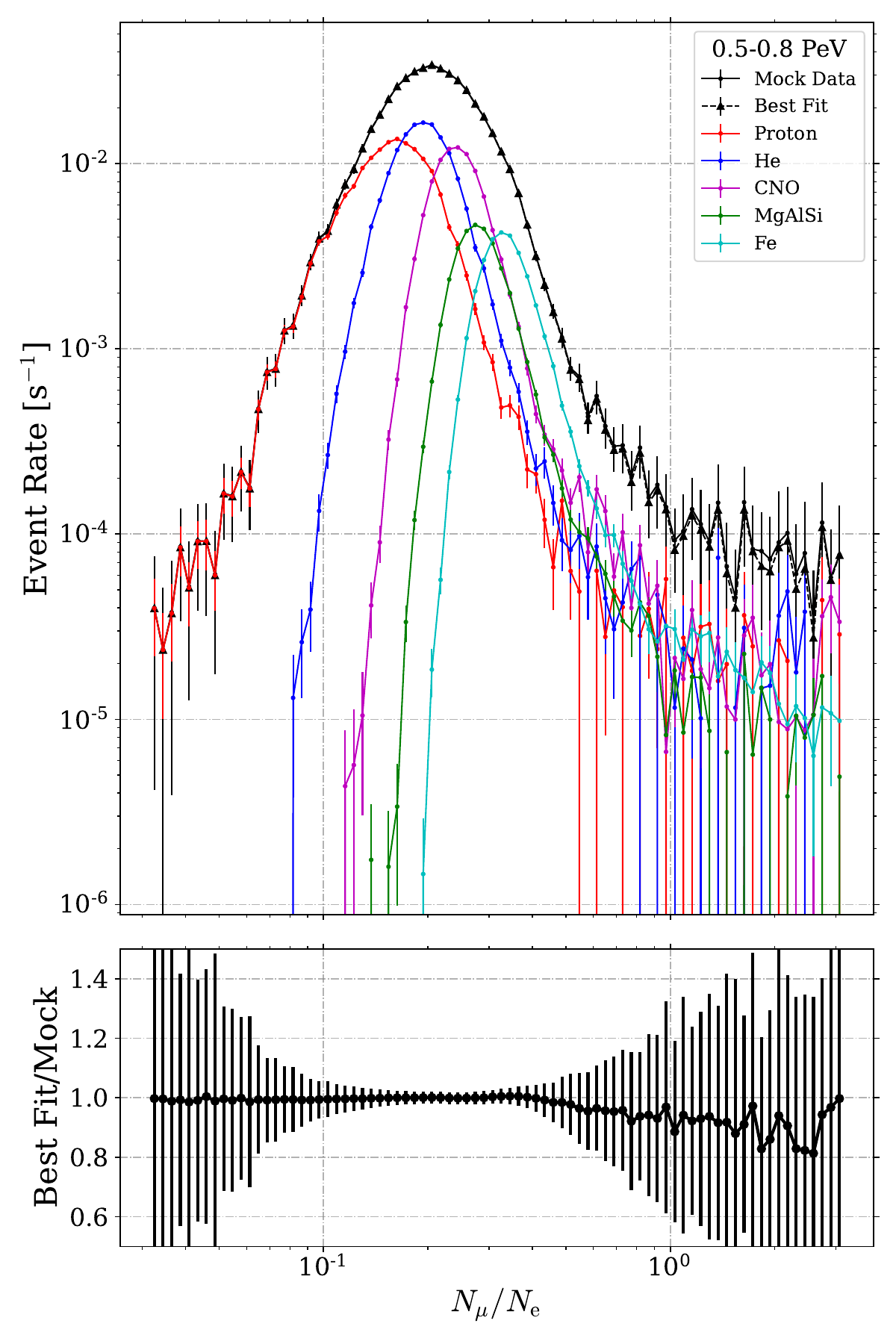}
            \caption{The best fit of \ratio distribution 
            in the energy bin of $0.5\, \textup{--}\, 0.8\si{PeV}$ (Test I).
            Upper panel: The mock data with GSF model (black solid line); the best fit (black dashed line); and the template distributions of individual mass groups with the H3a model: proton (red), He (blue), CNO (magenta), MgAlSi (green), and Fe (cyan). 
            Lower panel: the ratio between the best fit distribution
            and the mock data. The error bars are the statistical uncertainties due to limited simulated samples.} 
            \label{fig:fitted-ratio-spectrum}
        \end{figure}

        The flux of each mass group is fitted by maximizing
        the likelihood of observing the mock \ratio distribution.
        The template distributions of the five mass groups are
        adjusted vertically in the y-axis, i.e., fitting the fluxes of the mass groups,
        to find the parameters that maximize the likelihood.
        Figure \ref{fig:fitted-ratio-spectrum} shows, for example, a case of the best fit
        result for the \ratio distribution in
        the energy bin of {$0.5\,\text{--}\,0.8$ PeV}.
        The upper panel shows the \ratio
        distribution of the mock data and
        the best fit distribution, as well as the contributions from each mass group.
        The bottom panel shows the ratio between
        the best fit distribution and the mock data.
        One can see the best fit well restores the \ratio distribution. 
        
        It should be noted that in the MC test, the used template model, i.e.,
        the assumed spectral shape within an energy bin, is different from the input model for the mock observational data. The good recovery of the ratio distribution
        suggests the insensitivity of the fitting to the template model.
        
        However, the statistical uncertainty affects the uncertainties
        of the fitted fluxes, which is mainly from the limited MC sample size.
        In the energy bin of {$0.5\,\text{--}\,0.8 \si{PeV}$},
        the peak region of the \ratio distribution,
        with $0.1 < N_{\mu}/N_{\text{e}} < 0.3 $, is better
        sampled with smaller statistical uncertainties,
        whereas the side wings of the distribution have less
        statistics and thus larger fluctuation.
        For example, Fig. \ref{fig:fitted-ratio-spectrum} shows the right wing (\ratio$> 0.4$)
        is underestimated by the fitting,
        where the Fe group dominates the distribution.
        At higher energies, the fitting is more biased to the peak region
        of the \ratio distribution while the side wings are increasingly
        underestimated, compared to the mock data.
        
    \subsubsection{Reconstructed elemental energy spectra}
\begin{figure}
            \centering
            \includegraphics[width=3.4in]{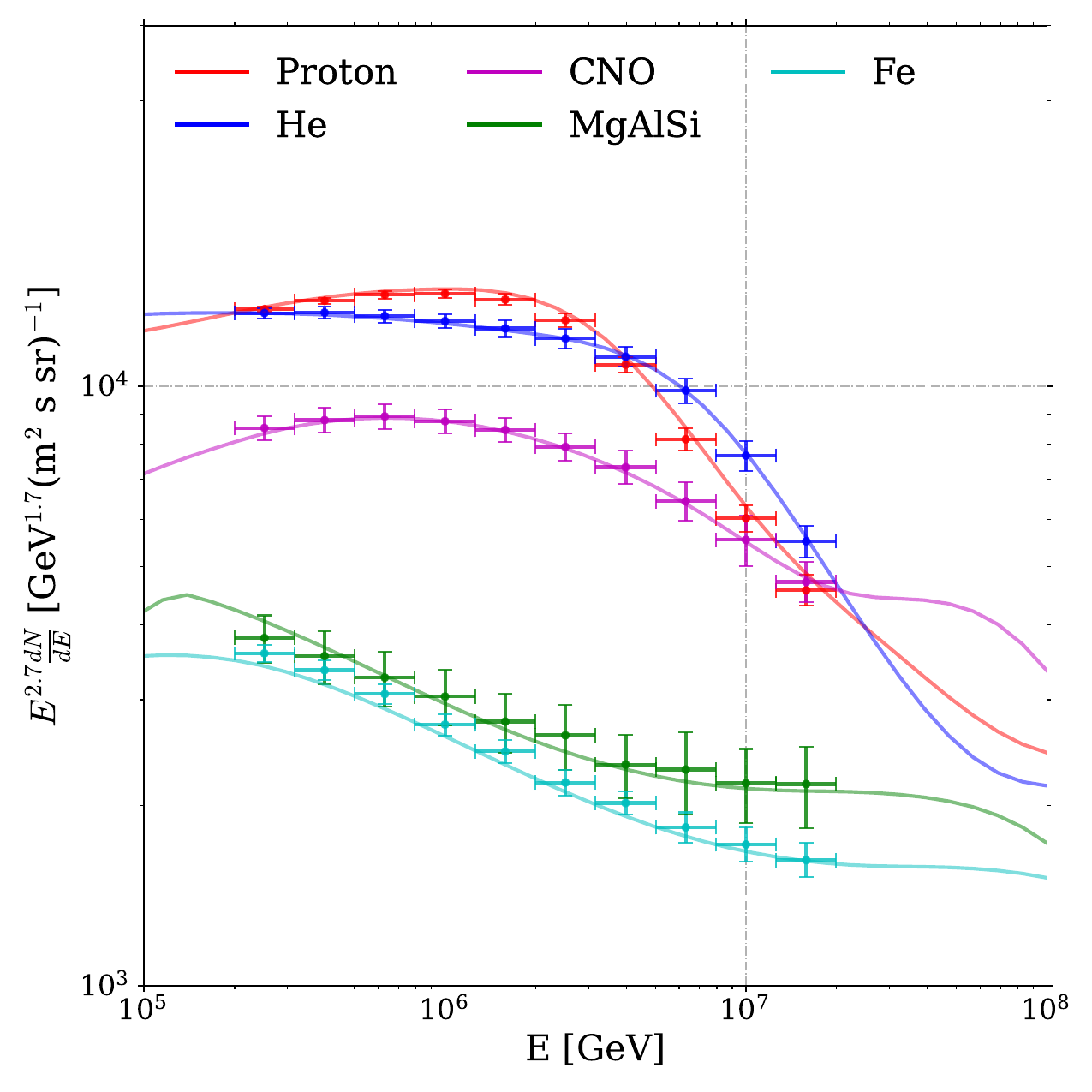}
            \caption{Reconstructed elemental energy spectra for GSF model in Test I. Different colors represent different mass groups,
        as annotated in the figure legend. The data points show the reconstructed elemental energy spectra of different mass groups, with the error bars given
        by the ${16\%}$ and ${84\%}$ percentiles of the posterior distributions. In comparison, the solid lines present the input GSF model.}
            \label{fig:rec-spectrum}
        \end{figure}

        The elemental energy spectra are reconstructed by
        fitting the \ratio distributions binned in primary energy.
        The reconstructed energy spectra of the five mass groups
        from $0.2\si{PeV}$ to $20\si{PeV}$
        are shown in Fig. \ref{fig:rec-spectrum},
        along with the input CR model, GSF, for comparison.       
        The derived fluxes are the median values of the a posteriori
        distributions sampled by MCMC.       
        The profiles of elemental energy spectra
        are well reconstructed in general. 
      
\begin{figure}
            \centering
            \includegraphics[width=3.4in]{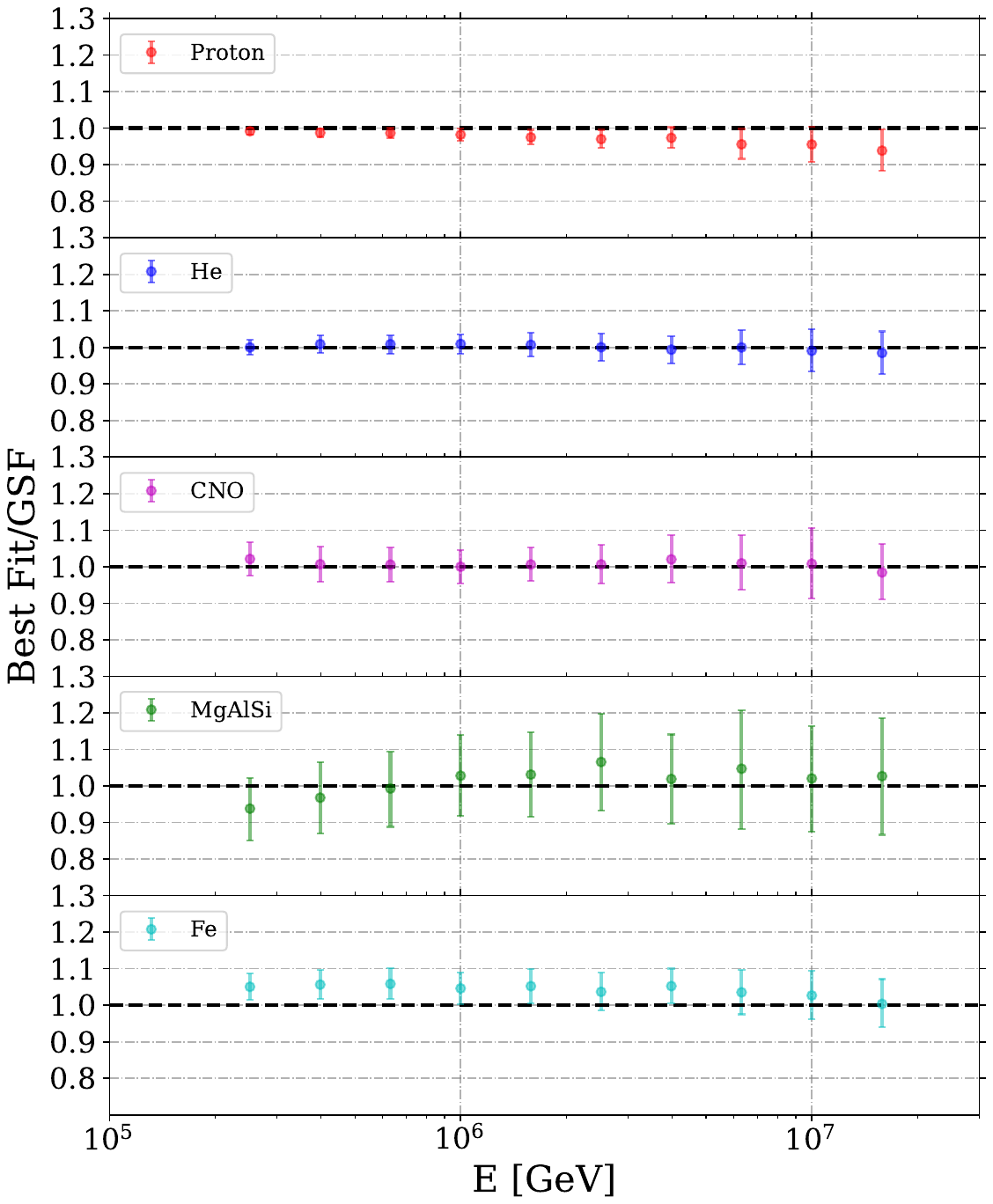}
            \caption{Ratios between reconstructed fluxes to the input ones as function of reconstructed energy in Test I.
            From top to bottom shown are for the five mass groups, proton, He, CNO, MgAlSi, and Fe.}
            \label{fig:ratio-GSF}
        \end{figure}

        The ratios of the best fit fluxes and the input model GSF
        are shown in Fig. \ref{fig:ratio-GSF} for the five mass groups.
        The GSF model is basically within the $1\sigma$ range of the reconstructed spectra.
        The reconstructed fluxes of all mass groups are well matching the input model.

        Overall, the bias of the reconstructed
        spectra is small, while the deviation also differs for mass groups. 
        The He and CNO spectra are best reconstructed, compared to the other mass groups, with the maximum deviations around $2\%$. The biases of the reconstructed energy spectra for proton and Fe are within $6\%$. And the reconstructed MgAlSi spectrum has the largest deviations from
        the GSF model, $\sim 7\%$.

        Given that the template and input models are different, the consistency between the reconstructed spectra and the mock data shows that the proposed approach works well to decompose the chemical composition of CRs around the knee.
                
        The anticorrelation between adjacent mass groups is also observed somewhat,
        e.g., the underestimate of CNO group in $0.5\, \text{--}\, 1 \si{PeV}$
        is correlated with the overestimate of MgAlSi in the same energy bin.
        The correlation and degeneracy between mass groups will be further discussed in
        Sec. \ref{sec:uncertainty}.
        
        

    \subsubsection{Uncertainties of reconstructed elemental energy spectra}\label{sec:uncertainty}
        The performance of the proposed CR decomposing method is further evaluated here by the uncertainties of the reconstructed elemental energy spectra. We find that: the uncertainties of the reconstructed spectra depend on both mass group and energy; and there is (anti-)correlation between the fitting results of different mass groups. 
        
\begin{figure}
            \centering
            \includegraphics[width=3.4in]{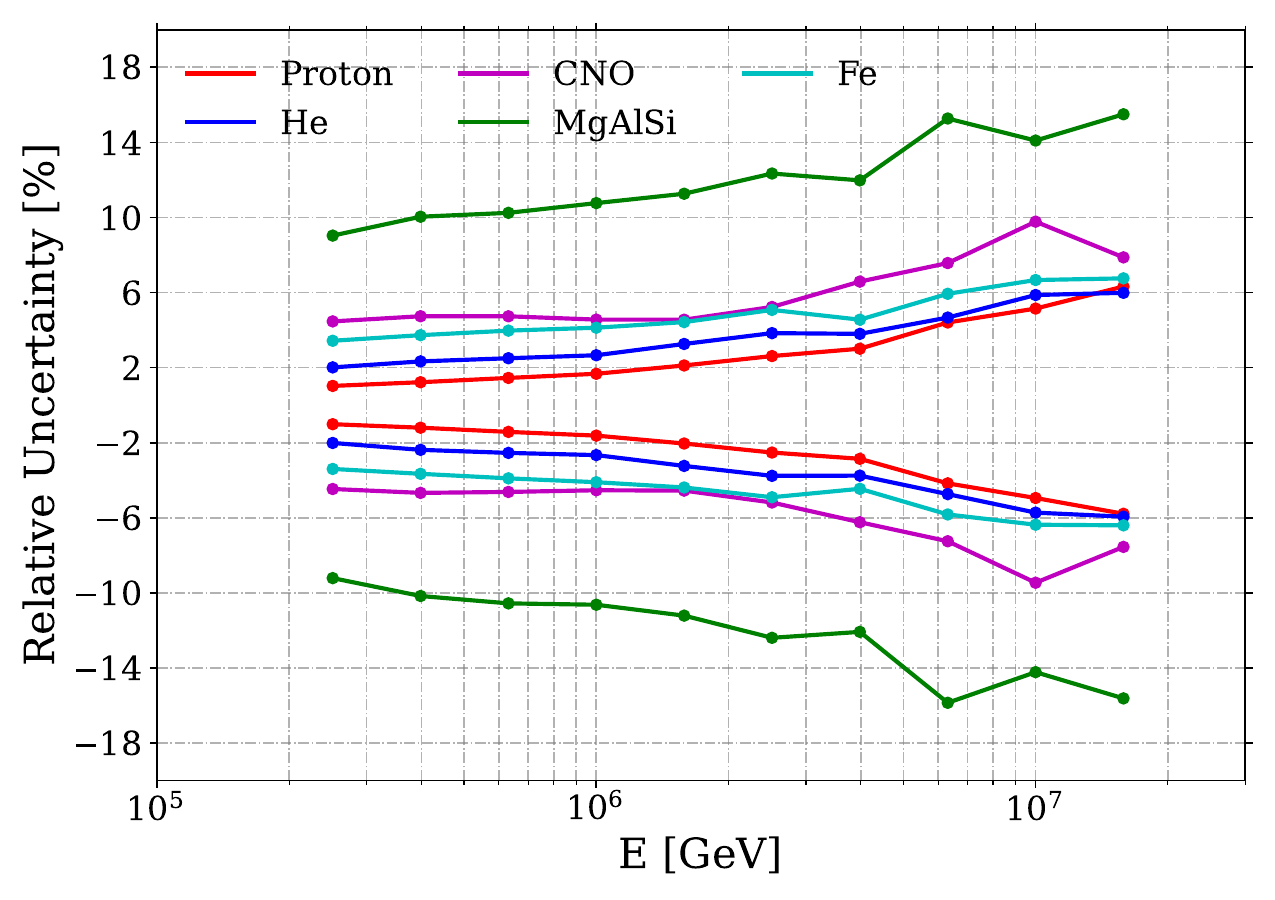}
            \caption{Relative uncertainties of the reconstructed elemental energy spectra as function of reconstructed energy in Test I. The colorful lines  correspond to different mass groups. }
            \label{fig:rel-err}
        \end{figure}

        As for the composition and energy dependence, one can see, from the fitting results, smaller uncertainty for proton,
        He, and Fe, but larger for CNO and MgAlSi; and larger uncertainty at higher energy. We show in Fig. \ref{fig:rel-err} the uncertainties of the best fit to the five mass groups in $0.2\, \text{--}\, 20 \si{PeV}$.
        For proton and He, the relative uncertainties 
        are {$2\%$} at $200\si{TeV}$ and {$6\%$} at $20\si{PeV}$;
        for Fe, the uncertainty increases from {$3\%$}
        at $200\si{TeV}$ to {$7\%$} at $20\si{PeV}$.
        As for CNO and MgAlSi, the uncertainties range
        from {$5\%$} and {$9\%$} at $200\si{TeV}$
        to {$8\%$} and {$16\%$} at $20\si{PeV}$, respectively.

        The reasons why proton, He and Fe groups are relatively better constrained could be understood. The \ratio distribution is usually a bump with two side wings dominated by the lightest and heaviest mass groups (Fig. \ref{fig:fitted-ratio-spectrum}), thus proton and Fe fluxes are constrained with smaller uncertainties. And He is next to proton in the distribution, with the \ratio peak well separated from the other groups, namely, proton and CNO (see Fig. \ref{fig:fitted-ratio-spectrum}), so the fitting of He flux is less affected by the mixing with proton and CNO groups.
        On the contrary, the two intermediate mass groups, CNO, and MgAlSi,
        which lie mostly around the peak region of the distribution,
        are subject to strong mixing of all the mass groups
        which makes the decomposition more challenging.
        Therefore, the determination of CNO and MgAlSi fluxes is less certain. 

        The reason why the fitting is better at low energies and more uncertain at high energies could be understood too.
        The dependence on energy of the uncertainty is mostly attributed
        to the limited MC sample size.
        Since the MC simulation is sampled from a $E^{-2}$ power-law,
        low energy showers are better sampled,
        while the high energy showers have smaller statistics
        with larger fluctuations.
        In principle, a larger simulation library can reduce
        the statistical uncertainties. However, the simulation of high
        energy shower is so time-consuming that the dependence
        on energy of the uncertainty is almost inevitable.
        
        \begin{figure}
            \centering
            \includegraphics[width=3.4in]{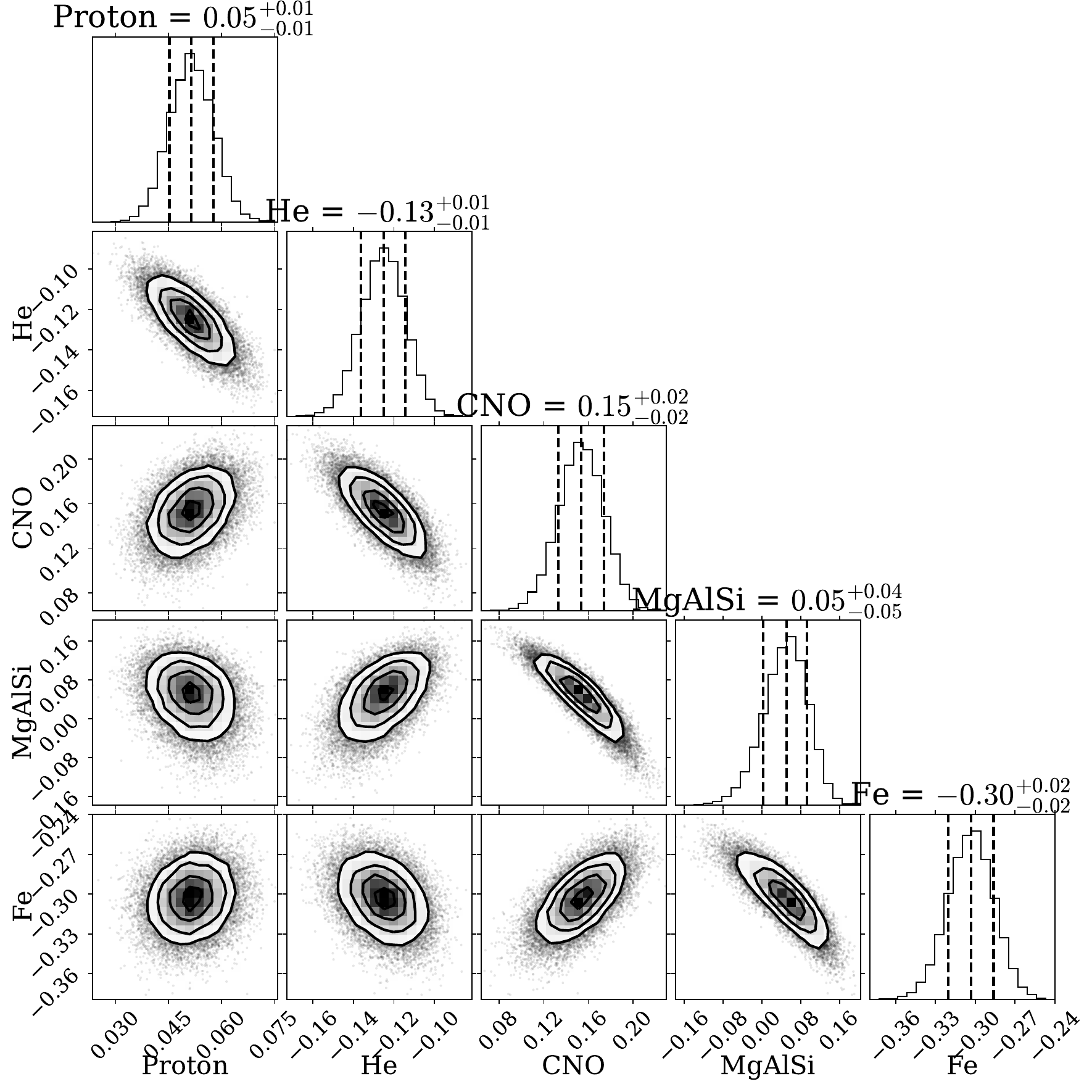}
            \caption{Posterior distributions and correlations of $\log f^{k}$ of
            the five mass groups in the energy bin of $0.5\,\text{--}\, 0.8 \si{PeV}$ (Test I).}
            \label{fig:corner}
        \end{figure}
        
        Finally, we find that there is correlation and degeneracy between fitting parameters, i.e., $f^k$ (with $k$ denoting different mass groups, see App. \ref{app:mcmc}),
        reflected by the posterior distributions of the MCMC.
        Figure \ref{fig:corner} shows, for example, the a posteriori distributions and the correlations
        between the fitting parameters in the energy bin of $0.5\,\text{--}\, 0.8 \si{PeV}$.
        The fitting parameters are $f^k$'s.
        The adjacent mass groups
        are anti-correlated, e.g., CNO and MgAlSi,
        which results from the overlap between the two adjacent mass groups.
        The mock data constrains the total number of events contributed
        by the two (or more) overlapping mass groups which leads to an anticorrelation or degeneracy
        between neighborhood nuclei. Therefore, the alternate mass groups
        are positively correlated, e.g., CNO and Fe.
        The correlation decreases for distant mass groups.
        For example, proton and Fe have the weakest correlation in Fig. \ref{fig:corner}, because they are best separated for the
        \ratio spectrum.
        The degeneracy between $f^k$'s is not unique to this analysis,
        but ubiquitous for any measurements on the elemental energy spectra.
        
    \subsection{Test for different template composition model}
    \label{sec:discussion:template model}
        Since the template distributions of mass groups depend on the 
        presumptions of spectral shapes within the energy bins, i.e., the template model, here we further carry out Test II, in comparison with Test I, to examine the dependence on template models. In Test II, we use the PG model to generate the template distributions, whereas the mock composition is kept the same as Test I.
        We approximate PG's mass groups of CNO, MgAlSi, and MnFeCo by simulating only the middle nucleus, i.e., N, Al, and Fe,
        and the nucleus flux is from summing fluxes
        of all three nuclei within the mass groups. 
        Template distributions of \ratio are produced with the PG model
        for the five mass groups after the above simplifications.
        
\begin{figure}
            \centering
            \includegraphics[width=3.4in]{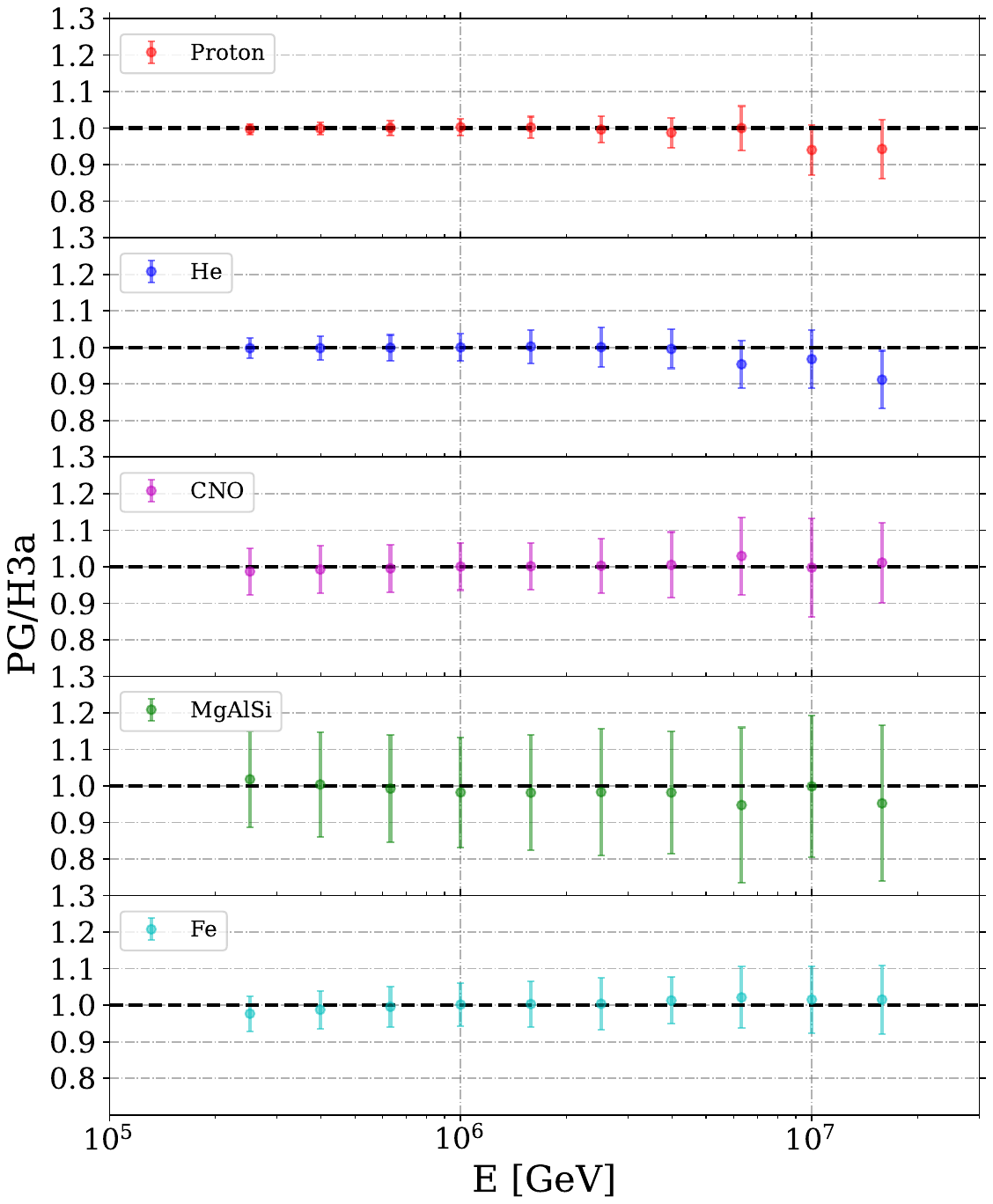}
            \caption{
            The ratio between the fluxes resulted from Test II and Test I (PG and H3a
            as template composition models, respectively) as function of the reconstructed energy. The error bar is only
            statistic uncertainty.
            }
            \label{fig:ratio-PG-template}
        \end{figure}

        Figure \ref{fig:ratio-PG-template} compares the results of Test II to Test I, under two different template models,
        i.e., PG versus H3a. The composition reconstruction is robust
        for changing the template models, where the difference between
        the two results is smaller than $9\%$. Particularly, the differences
        of results obtained with the two template models are
        within $2\%$ for low energies ($< 4 \si{PeV}$). The largest deviation is the He spectrum
        above $10\si{PeV}$ which likely comes from fluctuation caused by the small statistic
        of the MC sample.
        On the other hand, the errorbars represent the statistical
        uncertainties due to the limited simulation sample. 

        The consistency of the reconstructed spectra 
        with different template models demonstrates
        the decomposing method is almost independent on the CR composition model. This is because the MCMC fitting is done independently in each energy bin, and the energy bin is narrow enough so that the template spectral shape does not matter much.

    \subsection{Test for different input composition model}
    \label{sec:discussion:input model}
\begin{figure}
            \centering
            \includegraphics[width=3.4in]{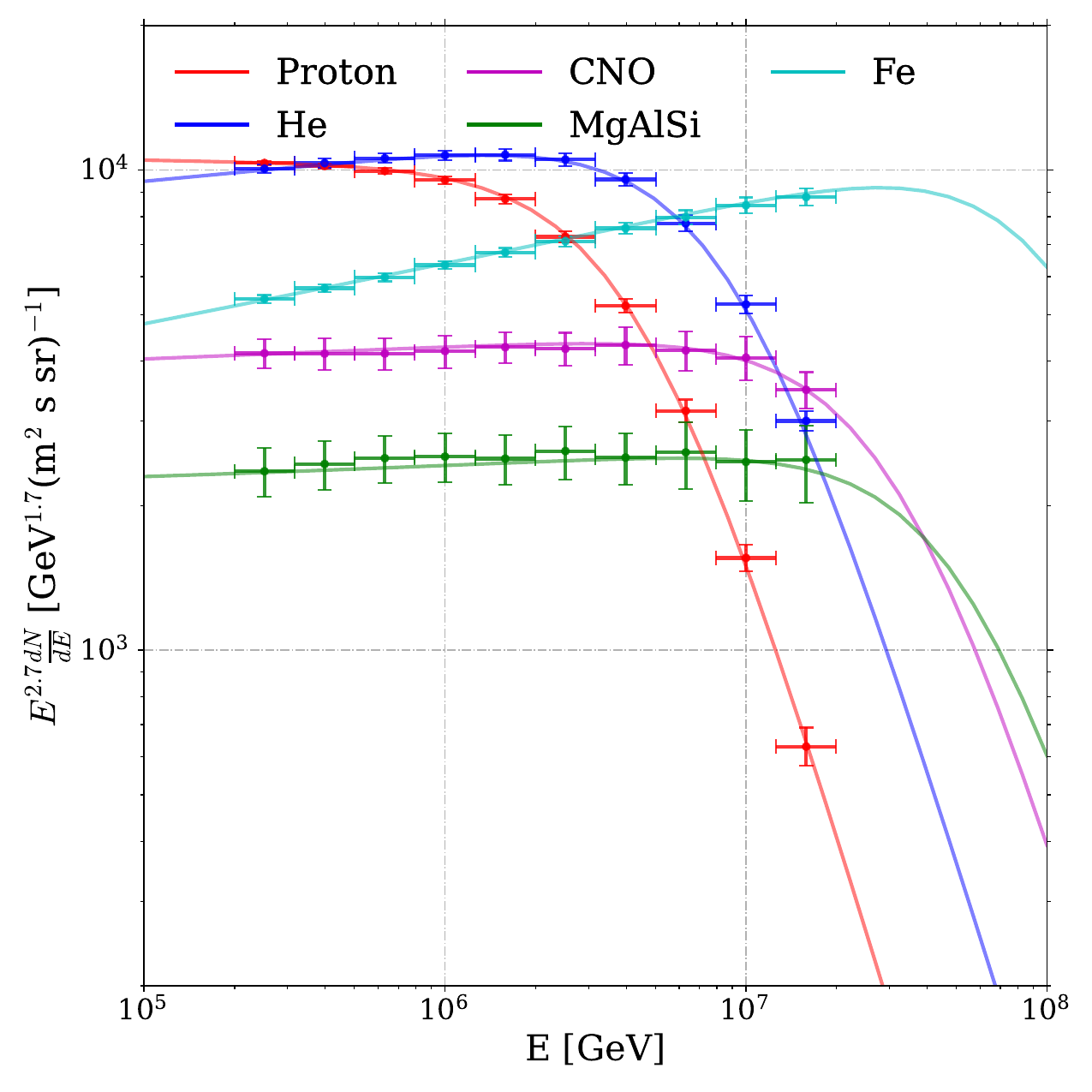}
            \caption{Same as Fig. \ref{fig:rec-spectrum}, but for Test III. 
            }
            \label{fig:spectra-PG}
        \end{figure}

\begin{figure}
            \centering
            \includegraphics[width=3.4in]{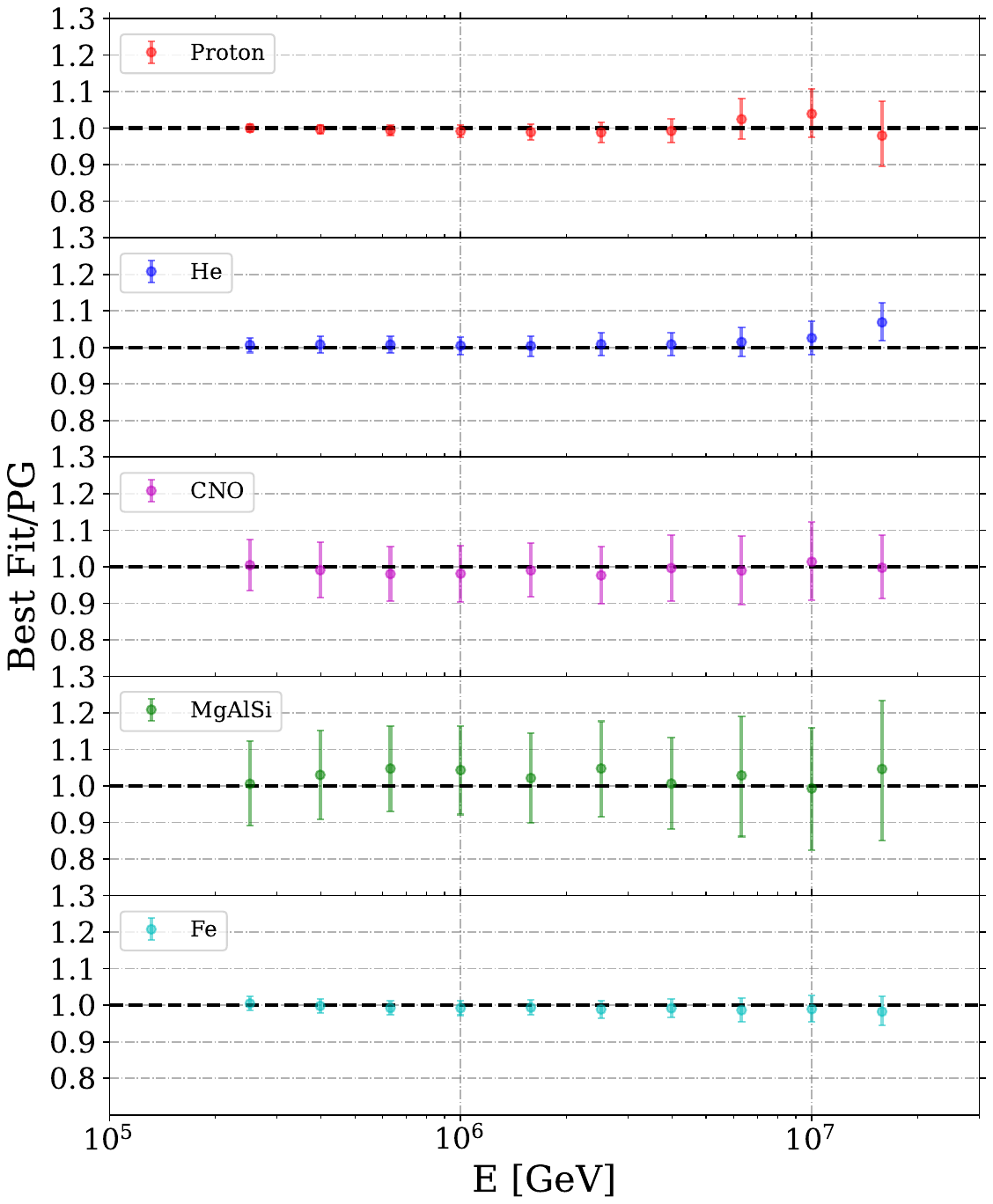}
            \caption{Same as Fig. \ref{fig:ratio-GSF}, but for Test III. 
              }
            \label{fig:ratio-PG}
        \end{figure}

        In order to test the reconstruction performance we also try the case of different input composition model. 
        Here we carry out Test III, where, compared to Test I, the input composition model is replaced by the PG model, whereas the the template model is kept the same.
        The reconstructed energy spectra for PG model are shown in Fig. \ref{fig:spectra-PG}.
        {The knee of proton and He below $10\si{PeV}$}
        are well restored by the reconstruction.
        The relative uncertainties are presented in Fig. \ref{fig:ratio-PG}.
        The reconstruction is done very well, with the reconstructed fluxes consistent with the PG model within the $1\sigma$ error. 
        
        The uncertainty changes very little compared with 
        the results of Test I. 
        For proton spectrum, the uncertainty
        is {$1\%$} at $200\si{TeV}$ and {$9\%$} at $20\si{PeV}$.
        For He and Fe groups, the uncertainties are
        {$2\%$} at $200\si{TeV}$ and up to {$5\%$} at $20\si{PeV}$.
        As for CNO and MgAlSi the uncertainty increases
        from {$7\%$} and {$12\%$} at $200\si{TeV}$
        to {$9\%$} and {$20\%$} at $20\si{PeV}$, respectively.
        
        The dependence of the uncertainties on the mass
        groups also exists for Test III.
        Again, the fluxes of proton,
        He, and Fe are better reconstructed,
        while the reconstruction of CNO and MgAlSi suffers more degeneracy due to
        stronger mixing of the two intermediate mass groups.
        
        The good recovery of the elemental energy spectra
        with a different input model demonstrates
        the reconstruction method is capable of restoring
        the CR composition in general, instead of 
        only for certain input models.
        
\section{Summary and discussion \label{sec:summary}}
    In this work, we propose an approach to reconstruct
    the elemental energy spectra of CRs around the knee with the \ratio distribution measured by LHAASO-KM2A: the energy spectra of individual CR compositions is obtained
    by fitting $N_{{\mu}} / N_{\mathrm{e}}$
    distributions in each reconstructed energy bin using the template shapes of $N_{{\mu}} / N_{\mathrm{e}}$ distributions of individual CR compositions based on MC simulation.
    We validate the feasibility of the approach
    and estimate the uncertainties of the reconstruction 
    with MC test, where we fit the mock data generated by MC simulation
    to test how well the input composition model can be restored.
    The uncertainties depend on mass and energy of the primary CR.
    In Test I using the GSF model as the input composition and the H3a model to
    generate template distributions, we find that the CR spectra of individual mass groups below 20 PeV can be well reconstructed by LHAASO-KM2A, for proton, He, and Fe groups, with uncertainty $<7\%$ below $20\si{PeV}$, though somewhat larger uncertainties for CNO and MgAlSi, $< 8\%$ and $< 16\%$ below $20\si{PeV}$, respectively.
    Furthermore, the versatility of the reconstruction is demonstrated by
    Test II and III where different template and input models are tested, showing good recovery of the input CR spectrum and weak dependence on the template model of individual mass groups. 
      
    The consistency  
        with different template models demonstrates
        the CR model independence of the method. This is mainly because the MCMC fitting is done independently between energy bins.
        This is especially desirable compared with those methods
        explicitly involving assumptions on CR composition, and avoids circular reasoning.
        Some methods using neutral network to decompose CR compositions
        necessitates a presumed composition model for training the network, e.g., 
        hypotheses for proton or heavy dominant \cite{RN546}, or equal mixture among four mass groups \cite{RN493}. However, the true CR composition distribution likely deviates from the assumption. 
    
        %
        
        We also caveat that the CR model independence 
        of this reconstruction method is at the expense of
        relying on an accurate description of \ratio distribution
        for the five mass groups.
        At low energies, MC simulation has sufficient
        statistic to sample the expected ratio distribution.
        At higher energies, the sampling may be incomplete
        due to poorer statistic, which introduces bias
        to the fitting.
        Considering the difficulty of simulation at high energies,
        parameterizing \ratio distribution from simulation may
        overcome the limited MC sample size with introducing 
        acceptable model dependence.
        For example, KASCADE used parameterization 
        to overcome the insufficiency of simulation
        and extrapolation to better describe the side wings
        of $N_{\mu}$ distribution that are hard to simulate \cite{RN544}.
        IceCube used kernel density estimation to extract 
        the template distributions of the four mass groups \cite{RN493}.
        Presently, the template distributions in this analysis are directly
        from the simulation data without parameterization,
        which is more based on simulation with minimized model dependence 
        on the choice of the parameterization.
  

In our method, the template distributions of individual mass groups rely on simulation, thus other uncertainties will arise when applying the method to measurements, i.e., the uncertainties from the hadronic interaction model and the atmosphere model, which is a general problem for decomposing method based on air shower simulation. {\color{black}In future analysis of measurements, more than one hadronic interaction model should be used to estimate the related systematic uncertainties. Moreover, the atmospheric profile could be more properly described. Also, the possible seasonal variation should be considered.
On the other hand, one may try to reconstruct the energy and composition together in order to mitigate the potential migration effect, especially for the events with small number of detectors triggered. }




\section*{Acknowledgment}
    This work is supported by the Natural Science Foundation of China (No. U1931201 and No. 12175121)
    and National Key R\&D Program of China No.2023YFE0102300.
    This work uses the {\small{crflux}} package \cite{PhysRevD.86.114024}
    to calculate the flux of CR models.
\appendix
\section{Event selection \label{app:cut}}
    Event selection filters out showers that are poorly
    reconstructed, e.g. showers with
    core position outside the array or showers
    with few secondary particles.
    On the other hand, the event selection depends
    on the scientific goal of the analysis to
    minimize the loss of events of interest.
    The selection conditions used in this analysis are as follows.
    (1) The zenith angle of the shower is from $10^{\circ}$
    to $30^{\circ}$, which is determined by matching
    the grammage of the atmosphere and the shower maximum
    of the energy range to be studied.
    (2) Distance of the shower core from the array center
    is from $320\ \si{m}$ to $420\ \si{m}$ to ensure the shower
    is well contained in the array with a good measurement
    of the em. and muonic content in the shower.
    (3) More than 50 EDs are triggered.
    (4) The number of secondary electromagnetic particles and muons
    located $40\, \text{--}\, 200\si{m}$ from the shower axis
    should be greater than $80$ and $15$, respectively.
    
    The geometry cut, i.e., cuts on the shower zenith angle and the shower core position,
    is used to select the showers that are well measured by KM2A array with little particles
    absorbed in the atmosphere or falling outside the array,
    which minimizes the bias and uncertainty in the reconstruction.
    The requirements on the minimum number of em. and muon particles is to guarantee the 
    reconstruction quality. The more secondary particles the array collects,
    better the reconstruction will be. At the same time,
    the minimum numbers of muons and electrons determine the threshold energy
    of the array, which makes the analysis start from $200\si{TeV}$.

\section{MCMC Fitting of \ratio distribution \label{app:mcmc}}
    The fluxes of the five mass groups are
    derived from fitting the \ratio distribution
    in the reconstructed energy bins.
    We use MCMC to sample
    the likelihood distribution of observing
    the actual data for different compositions,
    where the best fit maximizes the likelihood.

        The events are binned into primary energy $E$ and ratio $R\equiv N_\mu/N_{\rm e}$ bins. We define bin $(i,j)$ where $E_{i} < E < E_{i+1}$ and $R_{j} < R < R_{j+1}$.
        and $N^{\rm pred}_{i,j}$ and $N^{\rm obs}_{i,j}$ as the model-predicted and observed numbers of events, respectively, in bin $(i,j)$. 
    
        For a given composition model, the predicted number of events in a bin should be the sum of the contributions from all mass groups, and can be given by
            \begin{equation}
            N^{\text{pred}}_{i,j} = \sum_{k} f_i^{k} N_{i,j}^{\text{temp},k}
        \end{equation}
        where $N_{i,j}^{\text{temp},k}$ is the number of event
        contributed by nucleus $k$, with $k=$proton, He, CNO, MgAlSi, or Fe,
        in bin $(i, j)$ for
        the adopted template composition model, and $f_i^k$ is the 
        normalization factor for nucleus $k$ and energy bin $E_i$.
        $f_i^k$ is the parameter to be determined by comparing model with observation.
        
        We assume the number of events in a bin  
        can be approximated by a Gaussian distribution for large
        event counts. Thus, the likelihood in bin $(i, j)$ can be given by
        \begin{equation}
           \mathcal{L}_{i,j}(N^{\rm obs}_{i,j} | N^{\rm pred}_{i,j}) = \frac{1}{\sqrt{2 \pi }N^{\mathrm{pred}}_{i,j}}
            \exp\left[-\frac{(N^{\mathrm{obs}}_{i,j} - N^{\mathrm{pred}}_{i,j})^2}{2 N^{\mathrm{pred}}_{i,j}}\right],
        \end{equation}
        which depends on the fluxes of the five mass groups.
        In energy bin $i$, the likelihood
        considering all $R$ bins is
        \begin{equation}
            \mathcal{L}_{i}({N^{\mathrm{obs}}_{i}| N^{\mathrm{pred}}_{i}})
            \equiv \prod_{j} \mathcal{L}_{i,j}(N^{\rm obs}_{i,j} | N^{\rm pred}_{i,j}) ,
            \label{eq:likelihood}
        \end{equation}
        where the right hand side runs over all $R_j$ bins.
        Therefore, one can find the best-fit flux of each mass group at energy  bin $i$ by maximizing $\mathcal{L}_{i}({N^{\mathrm{obs}}_{i}| N^{\mathrm{pred}}_{i}})$.
        
        This can be done in each energy bin to derive $f_i^k$'s. 
        The best-fit flux can be given by $F^k(E_i) = f_i^k F_0^k(E_i)$, where $F_0^k(E_i)$ is the flux of nucleus $k$ at energy $E_i$ for the adopted template composition model.
        Therefore, the elemental energy spectra of each mass group are obtained.
        With $F_0^k(E_i)$ the flux of nucleus $k$ at
energy $E_i$ for the adopted template composition model,
the best-fit flux can be given by $F^k(E_i) = f_i^{k}F_0^k(E_i)$.
\bibliography{PeV_CR_10author}

\begin{thebibliography}{36}%
\makeatletter
\providecommand \@ifxundefined [1]{%
 \@ifx{#1\undefined}
}%
\providecommand \@ifnum [1]{%
 \ifnum #1\expandafter \@firstoftwo
 \else \expandafter \@secondoftwo
 \fi
}%
\providecommand \@ifx [1]{%
 \ifx #1\expandafter \@firstoftwo
 \else \expandafter \@secondoftwo
 \fi
}%
\providecommand \natexlab [1]{#1}%
\providecommand \enquote  [1]{``#1''}%
\providecommand \bibnamefont  [1]{#1}%
\providecommand \bibfnamefont [1]{#1}%
\providecommand \citenamefont [1]{#1}%
\providecommand \href@noop [0]{\@secondoftwo}%
\providecommand \href [0]{\begingroup \@sanitize@url \@href}%
\providecommand \@href[1]{\@@startlink{#1}\@@href}%
\providecommand \@@href[1]{\endgroup#1\@@endlink}%
\providecommand \@sanitize@url [0]{\catcode `\\12\catcode `\$12\catcode
  `\&12\catcode `\#12\catcode `\^12\catcode `\_12\catcode `\%12\relax}%
\providecommand \@@startlink[1]{}%
\providecommand \@@endlink[0]{}%
\providecommand \url  [0]{\begingroup\@sanitize@url \@url }%
\providecommand \@url [1]{\endgroup\@href {#1}{\urlprefix }}%
\providecommand \urlprefix  [0]{URL }%
\providecommand \Eprint [0]{\href }%
\providecommand \doibase [0]{https://doi.org/}%
\providecommand \selectlanguage [0]{\@gobble}%
\providecommand \bibinfo  [0]{\@secondoftwo}%
\providecommand \bibfield  [0]{\@secondoftwo}%
\providecommand \translation [1]{[#1]}%
\providecommand \BibitemOpen [0]{}%
\providecommand \bibitemStop [0]{}%
\providecommand \bibitemNoStop [0]{.\EOS\space}%
\providecommand \EOS [0]{\spacefactor3000\relax}%
\providecommand \BibitemShut  [1]{\csname bibitem#1\endcsname}%
\let\auto@bib@innerbib\@empty
\bibitem [{\citenamefont {Kulikov}\ and\ \citenamefont
  {Khristiansen}(1959)}]{RN588}%
  \BibitemOpen
  \bibfield  {author} {\bibinfo {author} {\bibfnamefont {G.}~\bibnamefont
  {Kulikov}}\ and\ \bibinfo {author} {\bibfnamefont {G.}~\bibnamefont
  {Khristiansen}},\ }\bibfield  {title} {\bibinfo {title} {{On the size
  spectrum of extensive air showers}},\ }\href@noop {} {\bibfield  {journal}
  {\bibinfo  {journal} {Sov. Phys. JETP}\ }\textbf {\bibinfo {volume} {35}},\
  \bibinfo {pages} {441} (\bibinfo {year} {1959})}\BibitemShut {NoStop}%
\bibitem [{\citenamefont {Stanev}\ \emph {et~al.}(1993)\citenamefont {Stanev},
  \citenamefont {Biermann},\ and\ \citenamefont {Gaisser}}]{RN576}%
  \BibitemOpen
  \bibfield  {author} {\bibinfo {author} {\bibfnamefont {T.}~\bibnamefont
  {Stanev}}, \bibinfo {author} {\bibfnamefont {P.~L.}\ \bibnamefont
  {Biermann}},\ and\ \bibinfo {author} {\bibfnamefont {T.~K.}\ \bibnamefont
  {Gaisser}},\ }\bibfield  {title} {\bibinfo {title} {{Cosmic rays. IV. The
  spectrum and chemical composition above $10^{4}$ GeV}},\ }\href
  {https://doi.org/10.48550/arXiv.astro-ph/9303006} {\bibfield  {journal}
  {\bibinfo  {journal} {Astron. and Astrophys.}\ }\textbf {\bibinfo {volume}
  {274}},\ \bibinfo {pages} {902} (\bibinfo {year} {1993})}\BibitemShut
  {NoStop}%
\bibitem [{\citenamefont {Kobayakawa}\ \emph {et~al.}(2002)\citenamefont
  {Kobayakawa}, \citenamefont {Honda},\ and\ \citenamefont {Samura}}]{RN577}%
  \BibitemOpen
  \bibfield  {author} {\bibinfo {author} {\bibfnamefont {K.}~\bibnamefont
  {Kobayakawa}}, \bibinfo {author} {\bibfnamefont {Y.~S.}\ \bibnamefont
  {Honda}},\ and\ \bibinfo {author} {\bibfnamefont {T.}~\bibnamefont
  {Samura}},\ }\bibfield  {title} {\bibinfo {title} {{Acceleration by oblique
  shocks at supernova remnants and cosmic ray spectra around the knee
  region}},\ }\href {https://doi.org/10.1103/PhysRevD.66.083004} {\bibfield
  {journal} {\bibinfo  {journal} {Phys. Rev. D}\ }\textbf {\bibinfo {volume}
  {66}},\ \bibinfo {pages} {083004} (\bibinfo {year} {2002})}\BibitemShut
  {NoStop}%
\bibitem [{\citenamefont {Baade}\ and\ \citenamefont {Zwicky}(1934)}]{RN575}%
  \BibitemOpen
  \bibfield  {author} {\bibinfo {author} {\bibfnamefont {W.}~\bibnamefont
  {Baade}}\ and\ \bibinfo {author} {\bibfnamefont {F.}~\bibnamefont {Zwicky}},\
  }\bibfield  {title} {\bibinfo {title} {{Remarks on Super-Novae and Cosmic
  Rays}},\ }\href {https://doi.org/10.1103/PhysRev.46.76.2} {\bibfield
  {journal} {\bibinfo  {journal} {Phys. Rev.}\ }\textbf {\bibinfo {volume}
  {46}},\ \bibinfo {pages} {76} (\bibinfo {year} {1934})}\BibitemShut {NoStop}%
\bibitem [{\citenamefont {Lagage}\ and\ \citenamefont
  {Cesarsky}(1983)}]{RN559}%
  \BibitemOpen
  \bibfield  {author} {\bibinfo {author} {\bibfnamefont {P.~O.}\ \bibnamefont
  {Lagage}}\ and\ \bibinfo {author} {\bibfnamefont {C.~J.}\ \bibnamefont
  {Cesarsky}},\ }\bibfield  {title} {\bibinfo {title} {{The maximum energy of
  cosmic rays accelerated by supernova shocks}},\ }\href
  {https://ui.adsabs.harvard.edu/abs/1983A&A...125..249L} {\bibfield  {journal}
  {\bibinfo  {journal} {Astron. and Astrophys.}\ }\textbf {\bibinfo {volume}
  {125}},\ \bibinfo {pages} {249} (\bibinfo {year} {1983})}\BibitemShut
  {NoStop}%
\bibitem [{\citenamefont {Gabici}\ \emph {et~al.}(2016)\citenamefont {Gabici},
  \citenamefont {Gaggero},\ and\ \citenamefont {Zandanel}}]{RN597}%
  \BibitemOpen
  \bibfield  {author} {\bibinfo {author} {\bibfnamefont {S.}~\bibnamefont
  {Gabici}}, \bibinfo {author} {\bibfnamefont {D.}~\bibnamefont {Gaggero}},\
  and\ \bibinfo {author} {\bibfnamefont {F.}~\bibnamefont {Zandanel}},\ }\href
  {https://doi.org/10.48550/arXiv.1610.07638} {\bibinfo {title} {{Can supernova
  remnants accelerate protons up to PeV energies?}}} (\bibinfo {year}
  {2016})\BibitemShut {NoStop}%
\bibitem [{\citenamefont {Ptuskin}\ \emph {et~al.}(1993)\citenamefont
  {Ptuskin}, \citenamefont {Rogovaya}, \citenamefont {Zirakashvili},
  \citenamefont {Chuvilgin}, \citenamefont {Khristiansen}, \citenamefont
  {Klepach},\ and\ \citenamefont {Kulikov}}]{RN585}%
  \BibitemOpen
  \bibfield  {author} {\bibinfo {author} {\bibfnamefont {V.~S.}\ \bibnamefont
  {Ptuskin}}, \bibinfo {author} {\bibfnamefont {S.~I.}\ \bibnamefont
  {Rogovaya}}, \bibinfo {author} {\bibfnamefont {V.~N.}\ \bibnamefont
  {Zirakashvili}}, \bibinfo {author} {\bibfnamefont {L.~G.}\ \bibnamefont
  {Chuvilgin}}, \bibinfo {author} {\bibfnamefont {G.~B.}\ \bibnamefont
  {Khristiansen}}, \bibinfo {author} {\bibfnamefont {E.~G.}\ \bibnamefont
  {Klepach}},\ and\ \bibinfo {author} {\bibfnamefont {G.~V.}\ \bibnamefont
  {Kulikov}},\ }\bibfield  {title} {\bibinfo {title} {{Diffusion and drift of
  very high energy cosmic rays in galactic magnetic fields}},\ }\href
  {https://ui.adsabs.harvard.edu/abs/1993A&A...268..726P} {\bibfield  {journal}
  {\bibinfo  {journal} {Astron. and Astrophys.}\ }\textbf {\bibinfo {volume}
  {68}},\ \bibinfo {pages} {726} (\bibinfo {year} {1993})}\BibitemShut
  {NoStop}%
\bibitem [{\citenamefont {Lagutin}\ \emph {et~al.}(2001)\citenamefont
  {Lagutin}, \citenamefont {Nikulin},\ and\ \citenamefont {Uchaikin}}]{RN582}%
  \BibitemOpen
  \bibfield  {author} {\bibinfo {author} {\bibfnamefont {A.~A.}\ \bibnamefont
  {Lagutin}}, \bibinfo {author} {\bibfnamefont {Y.~A.}\ \bibnamefont
  {Nikulin}},\ and\ \bibinfo {author} {\bibfnamefont {V.~V.}\ \bibnamefont
  {Uchaikin}},\ }\bibfield  {title} {\bibinfo {title} {{The “knee” in the
  primary cosmic ray spectrum as consequence of the anomalous diffusion of the
  particles in the fractal interstellar medium}},\ }\href
  {https://doi.org/https://doi.org/10.1016/S0920-5632(01)01280-4} {\bibfield
  {journal} {\bibinfo  {journal} {Nucl. Phys.}\ }\textbf {\bibinfo {volume}
  {97}},\ \bibinfo {pages} {267} (\bibinfo {year} {2001})}\BibitemShut
  {NoStop}%
\bibitem [{\citenamefont {Julián}\ \emph {et~al.}(2002)\citenamefont
  {Julián}, \citenamefont {Esteban},\ and\ \citenamefont {Luis}}]{RN578}%
  \BibitemOpen
  \bibfield  {author} {\bibinfo {author} {\bibfnamefont {C.}~\bibnamefont
  {Julián}}, \bibinfo {author} {\bibfnamefont {R.}~\bibnamefont {Esteban}},\
  and\ \bibinfo {author} {\bibfnamefont {N.~E.}\ \bibnamefont {Luis}},\
  }\bibfield  {title} {\bibinfo {title} {{Turbulent diffusion and drift in
  galactic magnetic fields and the explanation of the knee in the cosmic ray
  spectrum}},\ }\href {https://doi.org/10.1088/1126-6708/2002/12/033}
  {\bibfield  {journal} {\bibinfo  {journal} {J. High. Energy. Phys.}\ }\textbf
  {\bibinfo {volume} {2002}},\ \bibinfo {pages} {033} (\bibinfo {year}
  {2002})}\BibitemShut {NoStop}%
\bibitem [{\citenamefont {Juliusson}\ \emph {et~al.}(1972)\citenamefont
  {Juliusson}, \citenamefont {Meyer},\ and\ \citenamefont {Müller}}]{RN594}%
  \BibitemOpen
  \bibfield  {author} {\bibinfo {author} {\bibfnamefont {E.}~\bibnamefont
  {Juliusson}}, \bibinfo {author} {\bibfnamefont {P.}~\bibnamefont {Meyer}},\
  and\ \bibinfo {author} {\bibfnamefont {D.}~\bibnamefont {Müller}},\
  }\bibfield  {title} {\bibinfo {title} {{Composition of Cosmic-Ray Nuclei at
  High Energies}},\ }\href {https://doi.org/10.1103/PhysRevLett.29.445}
  {\bibfield  {journal} {\bibinfo  {journal} {Phys. Rev. Lett.}\ }\textbf
  {\bibinfo {volume} {29}},\ \bibinfo {pages} {445} (\bibinfo {year}
  {1972})}\BibitemShut {NoStop}%
\bibitem [{\citenamefont {Simpson}\ and\ \citenamefont
  {Garcia-Munoz}(1988)}]{RN595}%
  \BibitemOpen
  \bibfield  {author} {\bibinfo {author} {\bibfnamefont {J.~A.}\ \bibnamefont
  {Simpson}}\ and\ \bibinfo {author} {\bibfnamefont {M.}~\bibnamefont
  {Garcia-Munoz}},\ }\bibfield  {title} {\bibinfo {title} {{Cosmic-ray lifetime
  in the galaxy: Experimental results and models}},\ }\href
  {https://doi.org/10.1007/BF00212240} {\bibfield  {journal} {\bibinfo
  {journal} {Space. Sci. Rev.}\ }\textbf {\bibinfo {volume} {46}},\ \bibinfo
  {pages} {205} (\bibinfo {year} {1988})}\BibitemShut {NoStop}%
\bibitem [{\citenamefont {{B. Peters}}(1961)}]{RN433}%
  \BibitemOpen
  \bibfield  {author} {\bibinfo {author} {\bibnamefont {{B. Peters}}},\
  }\bibfield  {title} {\bibinfo {title} {{Primary cosmic radiation and
  extensive air showers}},\ }\href {https://doi.org/10.1007/BF02783106}
  {\bibfield  {journal} {\bibinfo  {journal} {Nuovo Cimento}\ }\textbf
  {\bibinfo {volume} {22}},\ \bibinfo {pages} {800} (\bibinfo {year}
  {1961})}\BibitemShut {NoStop}%
\bibitem [{\citenamefont {Gaisser}\ \emph {et~al.}(2013)\citenamefont
  {Gaisser}, \citenamefont {Stanev},\ and\ \citenamefont {Tilav}}]{RN426}%
  \BibitemOpen
  \bibfield  {author} {\bibinfo {author} {\bibfnamefont {T.~K.}\ \bibnamefont
  {Gaisser}}, \bibinfo {author} {\bibfnamefont {T.}~\bibnamefont {Stanev}},\
  and\ \bibinfo {author} {\bibfnamefont {S.}~\bibnamefont {Tilav}},\ }\bibfield
   {title} {\bibinfo {title} {{Cosmic ray energy spectrum from measurements of
  air showers}},\ }\href {https://doi.org/10.1007/s11467-013-0319-7} {\bibfield
   {journal} {\bibinfo  {journal} {Frontiers of Physics}\ }\textbf {\bibinfo
  {volume} {8}},\ \bibinfo {pages} {748} (\bibinfo {year} {2013})}\BibitemShut
  {NoStop}%
\bibitem [{\citenamefont {{P.A. Zyla \textit{et al.} (Particle Data
  Group)}}(2020)}]{RN557}%
  \BibitemOpen
  \bibfield  {author} {\bibinfo {author} {\bibnamefont {{P.A. Zyla \textit{et
  al.} (Particle Data Group)}}},\ }\bibfield  {title} {\bibinfo {title}
  {{Review of Particle Physics}},\ }\bibfield  {journal} {\bibinfo  {journal}
  {Prog. Theor. Exp. Phys.}\ }\textbf {\bibinfo {volume} {2020}},\ \href
  {https://doi.org/10.1093/ptep/ptaa104} {10.1093/ptep/ptaa104} (\bibinfo
  {year} {2020})\BibitemShut {NoStop}%
\bibitem [{\citenamefont {{T. Antoni, W. D. Apel, A. F. Badea, K. Bekk, A.
  Bercuci, J. Blümer, H. Bozdog, I. M. Brancus, A. Chilingarian, K. Daumiller
  \textit{et al.}}}(2005)}]{RN544}%
  \BibitemOpen
  \bibfield  {author} {\bibinfo {author} {\bibnamefont {{T. Antoni, W. D. Apel,
  A. F. Badea, K. Bekk, A. Bercuci, J. Blümer, H. Bozdog, I. M. Brancus, A.
  Chilingarian, K. Daumiller \textit{et al.}}}},\ }\bibfield  {title} {\bibinfo
  {title} {{KASCADE measurements of energy spectra for elemental groups of
  cosmic rays: Results and open problems}},\ }\href
  {https://doi.org/https://doi.org/10.1016/j.astropartphys.2005.04.001}
  {\bibfield  {journal} {\bibinfo  {journal} {Astropart. Phys.}\ }\textbf
  {\bibinfo {volume} {24}},\ \bibinfo {pages} {1} (\bibinfo {year}
  {2005})}\BibitemShut {NoStop}%
\bibitem [{\citenamefont {{M. Amenomori \textit{et al.} (Tibet AS$\gamma$
  Collaboration)}}(2006)}]{RN546}%
  \BibitemOpen
  \bibfield  {author} {\bibinfo {author} {\bibnamefont {{M. Amenomori
  \textit{et al.} (Tibet AS$\gamma$ Collaboration)}}},\ }\bibfield  {title}
  {\bibinfo {title} {{Are protons still dominant at the knee of the cosmic-ray
  energy spectrum?}},\ }\href
  {https://doi.org/https://doi.org/10.1016/j.physletb.2005.10.048} {\bibfield
  {journal} {\bibinfo  {journal} {Phys. Lett. B}\ }\textbf {\bibinfo {volume}
  {632}},\ \bibinfo {pages} {58} (\bibinfo {year} {2006})}\BibitemShut
  {NoStop}%
\bibitem [{\citenamefont {{B. Bartoli \textit{et al.} (ARGO-YBJ Collaboration,
  LHAASO Collaboration)}}(2015)}]{RN427}%
  \BibitemOpen
  \bibfield  {author} {\bibinfo {author} {\bibnamefont {{B. Bartoli \textit{et
  al.} (ARGO-YBJ Collaboration, LHAASO Collaboration)}}},\ }\bibfield  {title}
  {\bibinfo {title} {{Knee of the cosmic hydrogen and helium spectrum below 1
  PeV measured by ARGO-YBJ and a Cherenkov telescope of LHAASO}},\ }\href
  {https://doi.org/10.1103/PhysRevD.92.092005} {\bibfield  {journal} {\bibinfo
  {journal} {Phys. Rev. D}\ }\textbf {\bibinfo {volume} {92}},\ \bibinfo
  {pages} {092005} (\bibinfo {year} {2015})}\BibitemShut {NoStop}%
\bibitem [{\citenamefont {{Z. Cao, D. della Volpe, S. M. Liu, X. J. Bi, Y.
  Chen, B. D'Ettorre Piazzoli, L. Feng, H. Y Jia, Z. Li, X. H. Ma \textit{et
  al.}}}(2019)}]{RN590}%
  \BibitemOpen
  \bibfield  {author} {\bibinfo {author} {\bibnamefont {{Z. Cao, D. della
  Volpe, S. M. Liu, X. J. Bi, Y. Chen, B. D'Ettorre Piazzoli, L. Feng, H. Y
  Jia, Z. Li, X. H. Ma \textit{et al.}}}},\ }\bibfield  {title} {\bibinfo
  {title} {{The Large High Altitude Air Shower Observatory (LHAASO) Science
  Book (2021 Edition)}},\ }\bibfield  {journal} {\bibinfo  {journal} {Chin.
  Phys. C}\ }\textbf {\bibinfo {volume} {46}},\ \href
  {https://doi.org/10.48550/arXiv.1905.02773} {10.48550/arXiv.1905.02773}
  (\bibinfo {year} {2019})\BibitemShut {NoStop}%
\bibitem [{\citenamefont {{L. Q. Yin \textit{et al.} (LHAASO Collaboration)
  }}(2019)}]{RN623}%
  \BibitemOpen
  \bibfield  {author} {\bibinfo {author} {\bibnamefont {{L. Q. Yin \textit{et
  al.} (LHAASO Collaboration) }}},\ }\bibfield  {title} {\bibinfo {title}
  {{Expected energy spectrum of cosmic ray protons and helium below 4 PeV
  measured by LHAASO}},\ }\href {https://doi.org/10.1088/1674-1137/43/7/075001}
  {\bibfield  {journal} {\bibinfo  {journal} {Chin. Phys. C}\ }\textbf
  {\bibinfo {volume} {43}},\ \bibinfo {eid} {075001} (\bibinfo {year}
  {2019})}\BibitemShut {NoStop}%
\bibitem [{\citenamefont {{C. Jin, S. Z. Chen, and H. H. He (LHAASO
  Collaboration)}}(2020)}]{RN622}%
  \BibitemOpen
  \bibfield  {author} {\bibinfo {author} {\bibnamefont {{C. Jin, S. Z. Chen,
  and H. H. He (LHAASO Collaboration)}}},\ }\bibfield  {title} {\bibinfo
  {title} {{Classifying cosmic-ray proton and light groups in LHAASO-KM2A
  experiment with graph neural network}},\ }\href
  {https://doi.org/10.1088/1674-1137/44/6/065002} {\bibfield  {journal}
  {\bibinfo  {journal} {Chin. Phys. C}\ }\textbf {\bibinfo {volume} {44}},\
  \bibinfo {eid} {065002} (\bibinfo {year} {2020})}\BibitemShut {NoStop}%
\bibitem [{\citenamefont {{L. Q. Yin, S. H. Chen, and S. S. Zhang (LHAASO
  Collaboration)}}(2023)}]{RN624}%
  \BibitemOpen
  \bibfield  {author} {\bibinfo {author} {\bibnamefont {{L. Q. Yin, S. H. Chen,
  and S. S. Zhang (LHAASO Collaboration)}}},\ }\bibfield  {title} {\bibinfo
  {title} {{Measurement of the cosmic ray iron spectrum with energy from 100
  TeV to 10 PeV by LHAASO}},\ }\href {https://doi.org/10.22323/1.444.0419}
  {\bibfield  {journal} {\bibinfo  {journal} {PoS}\ }\textbf {\bibinfo {volume}
  {ICRC2023}},\ \bibinfo {pages} {419} (\bibinfo {year} {2023})}\BibitemShut
  {NoStop}%
\bibitem [{\citenamefont {Zhang}\ \emph {et~al.}(2022)\citenamefont {Zhang},
  \citenamefont {He},\ and\ \citenamefont {Feng}}]{RN542}%
  \BibitemOpen
  \bibfield  {author} {\bibinfo {author} {\bibfnamefont {H.~Y.}\ \bibnamefont
  {Zhang}}, \bibinfo {author} {\bibfnamefont {H.~H.}\ \bibnamefont {He}},\ and\
  \bibinfo {author} {\bibfnamefont {C.~F.}\ \bibnamefont {Feng}},\ }\bibfield
  {title} {\bibinfo {title} {{Approaches to composition independent energy
  reconstruction of cosmic rays based on the LHAASO-KM2A detector}},\ }\href
  {https://doi.org/10.1103/PhysRevD.106.123028} {\bibfield  {journal} {\bibinfo
   {journal} {Phys. Rev. D}\ }\textbf {\bibinfo {volume} {106}},\ \bibinfo
  {pages} {123028} (\bibinfo {year} {2022})}\BibitemShut {NoStop}%
\bibitem [{\citenamefont {{F. Aharonian \textit{et al.} (LHAASO
  Collaboration)}}(2021)}]{RN591}%
  \BibitemOpen
  \bibfield  {author} {\bibinfo {author} {\bibnamefont {{F. Aharonian
  \textit{et al.} (LHAASO Collaboration)}}},\ }\bibfield  {title} {\bibinfo
  {title} {{Observation of the Crab Nebula with LHAASO-KM2A − a performance
  study}},\ }\href {https://doi.org/10.1088/1674-1137/abd01b} {\bibfield
  {journal} {\bibinfo  {journal} {Chin. Phys. C}\ }\textbf {\bibinfo {volume}
  {45}},\ \bibinfo {pages} {025002} (\bibinfo {year} {2021})}\BibitemShut
  {NoStop}%
\bibitem [{\citenamefont {{X. H. Ma, Y. J. Bi, Z. Cao, M. J. Chen, S. Z. Chen,
  Y. D. Cheng, G. H. Gong, M. H. Gu, H. H. He, C. Hou \textit{et
  al.}}}(2022)}]{RN560}%
  \BibitemOpen
  \bibfield  {author} {\bibinfo {author} {\bibnamefont {{X. H. Ma, Y. J. Bi, Z.
  Cao, M. J. Chen, S. Z. Chen, Y. D. Cheng, G. H. Gong, M. H. Gu, H. H. He, C.
  Hou \textit{et al.}}}},\ }\bibfield  {title} {\bibinfo {title} {{Chapter 1
  LHAASO Instruments and Detector technology}},\ }\href
  {https://doi.org/10.1088/1674-1137/ac3fa6} {\bibfield  {journal} {\bibinfo
  {journal} {Chin. Phys. C}\ }\textbf {\bibinfo {volume} {46}},\ \bibinfo
  {pages} {030001} (\bibinfo {year} {2022})}\BibitemShut {NoStop}%
\bibitem [{\citenamefont {Heck}\ \emph {et~al.}(1998)\citenamefont {Heck},
  \citenamefont {Knapp}, \citenamefont {Capdevielle}, \citenamefont {Schatz},\
  and\ \citenamefont {Thouw}}]{RN553}%
  \BibitemOpen
  \bibfield  {author} {\bibinfo {author} {\bibfnamefont {D.}~\bibnamefont
  {Heck}}, \bibinfo {author} {\bibfnamefont {J.}~\bibnamefont {Knapp}},
  \bibinfo {author} {\bibfnamefont {J.~N.}\ \bibnamefont {Capdevielle}},
  \bibinfo {author} {\bibfnamefont {G.}~\bibnamefont {Schatz}},\ and\ \bibinfo
  {author} {\bibfnamefont {T.}~\bibnamefont {Thouw}},\ }\href
  {https://ui.adsabs.harvard.edu/abs/1998cmcc.book.....H} {\emph {\bibinfo
  {title} {{\small CORSIKA}: a Monte Carlo code to simulate extensive air
  showers}}}\ (\bibinfo {year} {1998})\BibitemShut {NoStop}%
\bibitem [{\citenamefont {Chen}\ \emph {et~al.}(2017)\citenamefont {Chen},
  \citenamefont {Zhao},\ and\ \citenamefont {Liu}}]{RN570}%
  \BibitemOpen
  \bibfield  {author} {\bibinfo {author} {\bibfnamefont {S.~Z.}\ \bibnamefont
  {Chen}}, \bibinfo {author} {\bibfnamefont {J.}~\bibnamefont {Zhao}},\ and\
  \bibinfo {author} {\bibfnamefont {Y.}~\bibnamefont {Liu}},\ }\bibfield
  {title} {\bibinfo {title} {{Design of the detector simulation of
  LHAASO-KM2A}},\ }\href@noop {} {\bibfield  {journal} {\bibinfo  {journal}
  {Nucl. Electron. Detect. Technol.}\ }\textbf {\bibinfo {volume} {37}},\
  \bibinfo {pages} {1101} (\bibinfo {year} {2017})}\BibitemShut {NoStop}%
\bibitem [{\citenamefont {Ostapchenko}(2013)}]{RN563}%
  \BibitemOpen
  \bibfield  {author} {\bibinfo {author} {\bibfnamefont {S.}~\bibnamefont
  {Ostapchenko}},\ }\bibfield  {title} {\bibinfo {title} {{QGSJET-II: physics,
  recent improvements, and results for air showers}},\ }\href
  {https://doi.org/10.1051/epjconf/20125202001} {\bibfield  {journal} {\bibinfo
   {journal} {EPJ Web of Conferences}\ }\textbf {\bibinfo {volume} {52}},\
  \bibinfo {pages} {02001} (\bibinfo {year} {2013})}\BibitemShut {NoStop}%
\bibitem [{\citenamefont {{G. Battistoni, T. Boehlen, F. Cerutti, P. W. Chin,
  L. S. Esposito, A. Fassò, A. Ferrari, A. Lechner, A. Empl, A. Mairani
  \textit{et al.}}}(2015)}]{RN564}%
  \BibitemOpen
  \bibfield  {author} {\bibinfo {author} {\bibnamefont {{G. Battistoni, T.
  Boehlen, F. Cerutti, P. W. Chin, L. S. Esposito, A. Fassò, A. Ferrari, A.
  Lechner, A. Empl, A. Mairani \textit{et al.}}}},\ }\bibfield  {title}
  {\bibinfo {title} {{Overview of the {\small FLUKA} code}},\ }\href
  {https://doi.org/https://doi.org/10.1016/j.anucene.2014.11.007} {\bibfield
  {journal} {\bibinfo  {journal} {Ann. Nucl. Energy}\ }\textbf {\bibinfo
  {volume} {82}},\ \bibinfo {pages} {10} (\bibinfo {year} {2015})}\BibitemShut
  {NoStop}%
\bibitem [{\citenamefont {Matthews}(2005)}]{RN297}%
  \BibitemOpen
  \bibfield  {author} {\bibinfo {author} {\bibfnamefont {J.}~\bibnamefont
  {Matthews}},\ }\bibfield  {title} {\bibinfo {title} {{A Heitler model of
  extensive air showers}},\ }\href
  {https://doi.org/https://doi.org/10.1016/j.astropartphys.2004.09.003}
  {\bibfield  {journal} {\bibinfo  {journal} {Astropart. Phys.}\ }\textbf
  {\bibinfo {volume} {22}},\ \bibinfo {pages} {387} (\bibinfo {year}
  {2005})}\BibitemShut {NoStop}%
\bibitem [{\citenamefont {Hörandel}(2007)}]{RN561}%
  \BibitemOpen
  \bibfield  {author} {\bibinfo {author} {\bibfnamefont {J.~R.}\ \bibnamefont
  {Hörandel}},\ }\bibfield  {title} {\bibinfo {title} {{Cosmic Rays from the
  Knee to the Second Knee:. $10^{14}$ to $10^{18}$eV}},\ }\href
  {https://doi.org/10.1142/s0217732307024139} {\bibfield  {journal} {\bibinfo
  {journal} {Mod. Phys. Lett. A}\ }\textbf {\bibinfo {volume} {22}},\ \bibinfo
  {pages} {1533} (\bibinfo {year} {2007})}\BibitemShut {NoStop}%
\bibitem [{\citenamefont {Gaisser}(2012)}]{RN261}%
  \BibitemOpen
  \bibfield  {author} {\bibinfo {author} {\bibfnamefont {T.~K.}\ \bibnamefont
  {Gaisser}},\ }\bibfield  {title} {\bibinfo {title} {{Spectrum of cosmic-ray
  nucleons, kaon production, and the atmospheric muon charge ratio}},\ }\href
  {https://doi.org/https://doi.org/10.1016/j.astropartphys.2012.02.010}
  {\bibfield  {journal} {\bibinfo  {journal} {Astropart. Phys.}\ }\textbf
  {\bibinfo {volume} {35}},\ \bibinfo {pages} {801} (\bibinfo {year}
  {2012})}\BibitemShut {NoStop}%
\bibitem [{\citenamefont {Foreman-Mackey}\ \emph {et~al.}(2013)\citenamefont
  {Foreman-Mackey}, \citenamefont {Hogg}, \citenamefont {Lang},\ and\
  \citenamefont {Goodman}}]{RN572}%
  \BibitemOpen
  \bibfield  {author} {\bibinfo {author} {\bibfnamefont {D.}~\bibnamefont
  {Foreman-Mackey}}, \bibinfo {author} {\bibfnamefont {D.~W.}\ \bibnamefont
  {Hogg}}, \bibinfo {author} {\bibfnamefont {D.}~\bibnamefont {Lang}},\ and\
  \bibinfo {author} {\bibfnamefont {J.}~\bibnamefont {Goodman}},\ }\bibfield
  {title} {\bibinfo {title} {{emcee: The MCMC Hammer}},\ }\href
  {https://doi.org/10.1086/670067} {\bibfield  {journal} {\bibinfo  {journal}
  {Publ. Astron. Soc. Pac.}\ }\textbf {\bibinfo {volume} {125}},\ \bibinfo
  {pages} {306} (\bibinfo {year} {2013})}\BibitemShut {NoStop}%
\bibitem [{\citenamefont {Dembinski}\ \emph {et~al.}(2017)\citenamefont
  {Dembinski}, \citenamefont {Engel}, \citenamefont {Fedynitch}, \citenamefont
  {Gaisser}, \citenamefont {Riehn},\ and\ \citenamefont {Stanev}}]{RN569}%
  \BibitemOpen
  \bibfield  {author} {\bibinfo {author} {\bibfnamefont {H.}~\bibnamefont
  {Dembinski}}, \bibinfo {author} {\bibfnamefont {R.}~\bibnamefont {Engel}},
  \bibinfo {author} {\bibfnamefont {A.}~\bibnamefont {Fedynitch}}, \bibinfo
  {author} {\bibfnamefont {T.}~\bibnamefont {Gaisser}}, \bibinfo {author}
  {\bibfnamefont {F.}~\bibnamefont {Riehn}},\ and\ \bibinfo {author}
  {\bibfnamefont {T.}~\bibnamefont {Stanev}},\ }\bibfield  {title} {\bibinfo
  {title} {{Data-driven model of the cosmic-ray flux and mass composition from
  10 GeV to $10^{11}$ GeV}},\ }\href {https://doi.org/10.22323/1.301.0533}
  {\bibfield  {journal} {\bibinfo  {journal} {Proc. Sci.}\ }\textbf {\bibinfo
  {volume} {ICRC2017}},\ \bibinfo {pages} {533} (\bibinfo {year}
  {2017})}\BibitemShut {NoStop}%
\bibitem [{\citenamefont {Hörandel}(2003)}]{RN423}%
  \BibitemOpen
  \bibfield  {author} {\bibinfo {author} {\bibfnamefont {J.~R.}\ \bibnamefont
  {Hörandel}},\ }\bibfield  {title} {\bibinfo {title} {{On the knee in the
  energy spectrum of cosmic rays}},\ }\href
  {https://doi.org/10.1016/s0927-6505(02)00198-6} {\bibfield  {journal}
  {\bibinfo  {journal} {Astropart. Phys.}\ }\textbf {\bibinfo {volume} {19}},\
  \bibinfo {pages} {193} (\bibinfo {year} {2003})}\BibitemShut {NoStop}%
\bibitem [{\citenamefont {{M. G. Aartsen \textit{et al.} (IceCube
  Collaboration)}}(2019)}]{RN493}%
  \BibitemOpen
  \bibfield  {author} {\bibinfo {author} {\bibnamefont {{M. G. Aartsen
  \textit{et al.} (IceCube Collaboration)}}},\ }\bibfield  {title} {\bibinfo
  {title} {{Cosmic ray spectrum and composition from PeV to EeV using 3 years
  of data from IceTop and IceCube}},\ }\href
  {https://doi.org/10.1103/PhysRevD.100.082002} {\bibfield  {journal} {\bibinfo
   {journal} {Phys. Rev. D}\ }\textbf {\bibinfo {volume} {100}},\ \bibinfo
  {pages} {082002} (\bibinfo {year} {2019})}\BibitemShut {NoStop}%
\bibitem [{\citenamefont {Fedynitch}\ \emph {et~al.}(2012)\citenamefont
  {Fedynitch}, \citenamefont {Becker~Tjus},\ and\ \citenamefont
  {Desiati}}]{PhysRevD.86.114024}%
  \BibitemOpen
  \bibfield  {author} {\bibinfo {author} {\bibfnamefont {A.}~\bibnamefont
  {Fedynitch}}, \bibinfo {author} {\bibfnamefont {J.}~\bibnamefont
  {Becker~Tjus}},\ and\ \bibinfo {author} {\bibfnamefont {P.}~\bibnamefont
  {Desiati}},\ }\bibfield  {title} {\bibinfo {title} {Influence of hadronic
  interaction models and the cosmic ray spectrum on the high energy atmospheric
  muon and neutrino flux},\ }\href {https://doi.org/10.1103/PhysRevD.86.114024}
  {\bibfield  {journal} {\bibinfo  {journal} {Phys. Rev. D}\ }\textbf {\bibinfo
  {volume} {86}},\ \bibinfo {pages} {114024} (\bibinfo {year}
  {2012})}\BibitemShut {NoStop}%
\end{thebibliography}%
\end{document}